%% file: MainPaper.tex
\documentclass[manuscript]{acmart}
\AtBeginDocument{%
  }

\acmJournal{TOSEM}
\acmYear{2026}
\acmDOI{}
\usepackage[table]{xcolor}  
\usepackage[utf8]{inputenc}
\usepackage{tabularx} 
\usepackage{array}
\usepackage{multirow}
\usepackage{float}
\usepackage{makecell}

\usepackage{amsmath,amssymb}
\usepackage{soul}
\usepackage{fontawesome5}
\usepackage{xspace}
\usepackage{subcaption}
\usepackage{calc}  
\usepackage{listings}

\lstdefinestyle{codecell}{
    basicstyle=\ttfamily\scriptsize,
    breaklines=true,
    columns=fullflexible,
    frame=none,
    xleftmargin=0pt,
    xrightmargin=0pt,
    aboveskip=0pt,
    belowskip=0pt
}

\newcommand{\CMone}{Brainstormer\xspace}
\newcommand{\CMtwo}{Implementer\xspace}
\newcommand{\CMthree}{Verifier\xspace}
\newcommand{\CMfour}{Copilot\xspace}
\newcommand{\Atask}{Algorithmic tasks\xspace}
\newcommand{\Btask}{System Design tasks\xspace}

\definecolor{ctrlBar}{RGB}{31,119,180}  
\definecolor{treatBar}{RGB}{255,127,14}

\newcommand{\CtrlBar}[1]{{\color{ctrlBar}\rule{\dimexpr 0.025cm*#1\relax}{0.2cm}}\hspace{3pt}\small #1\%}
\newcommand{\TreatBar}[1]{{\color{treatBar}\rule{\dimexpr 0.025cm*#1\relax}{0.2cm}}\hspace{3pt}\small #1\%}
  
\newcommand{\inlinequote}[1]{{``\textit{#1}''}}

\newcommand{\blockquote}[1]{%
  \par\vspace{0.2em}
  {\list{}{\leftmargin=1em \rightmargin=1em \parsep=0pt}%
   \item \emph{``#1''}%
   \endlist}%
  \vspace{0.2em}
}

\newif\ifdraft     
\draftfalse       

\newcommand{\modified}[1]{%
  \ifdraft
    \textcolor{black}{#1}%
  \else
    #1%
  \fi
}

\newcommand{\revised}[1]{%
  \ifdraft
    \textcolor{black}{#1}%
  \else
    #1%
  \fi
}

\newcommand{\rrevised}[1]{%
  \ifdraft
    \textcolor{black}{#1}%
  \else
    #1%
  \fi
}
\newcommand{\new}[1]{#1}

\begin{document}

\title[Exploring the Impact of LLM Interaction on the Creative Process of Programming]{
``Like Taking the Path of Least Resistance'': Exploring the Impact of LLM Interaction on the Creative Process of Programming}

\author{Zeinabsadat Saghi}
\orcid{0009-0007-3414-5688}
\email{saghi@usc.edu}
\affiliation{%
  \institution{Thomas Lord Department of Computer Science, University of Southern California}
  \city{Los Angeles}
  \state{CA}
  \country{USA}
}

\author{Run Huang}
\orcid{https://orcid.org/0009-0009-7467-4896}
\affiliation{%
  \institution{Thomas Lord Department of Computer Science, University of Southern California}
  \city{Los Angeles}
  \state{CA}
  \country{USA}
}
\email{runhuang@usc.edu}

\author{Souti Chattopadhyay}
\orcid{https://orcid.org/0000-0003-1644-7344}
\affiliation{%
  \institution{Thomas Lord Department of Computer Science, University of Southern California}
  \city{Los Angeles}
  \state{CA}
  \country{USA}
}
\email{schattop@usc.edu}

\renewcommand{\shortauthors}{Saghi et al.}

\begin{abstract}

\input{Sections/abstract}
\end{abstract}



\maketitle

\input{Sections/Intro}

\input{Sections/RW}
\input{Sections/Method}
\input{Sections/Analysis}

\input{Sections/RQ1}

\input{Sections/RQ2}

\input{Sections/RQ3}
\input{Sections/Discussion}
\input{Sections/Limitations}

\input{Sections/Conclusion}

\bibliographystyle{ACM-Reference-Format}
\bibliography{Sections/References}

\end{document}

%% file: Sections/abstract.tex
As AI takes on more of the generative work that once required human imagination, despite documented limitations in creative ability, a critical question emerges: \textit{How does GenAI affect users' creativity?} Through a within-subject study followed by retrospective interviews with (N=20) programmers, we investigated the impact of LLMs on participants' process of creative thinking in programming and the creativity of generated solutions. Across two conditions (LLM-assisted vs. unassisted), participants using LLMs had significantly shorter idea-generation periods (\rrevised{$p=1.6\times10^{-5}$}) and \rrevised{fewer creative moments ($p=4.4\times10^{-7}$)}. Qualitative analysis of participants' interactions and interviews revealed four different human-LLM collaboration modes supporting various problem-solving strategies.
\rrevised{However, a comparative analysis of the generated solutions shows that while LLMs can help generate more correct and functional code, LLM-assisted solutions to algorithmic tasks were significantly more similar to one another, even though the number of unique ideas was comparable.}
Based on our findings, we discuss design implications and considerations for effectively using LLMs to support user creativity.

%% file: Sections/Intro.tex
\section{Introduction}

While increasingly powerful AI systems promise unprecedented productivity gains, emerging research reveals a critical threat: users systematically over-rely on LLMs, relinquishing their own creativity and thinking processes to AI-generated solutions~\cite{lee2022coauthor, yuan2022wordcraft}. As one study participant explained, the decision to delegate creative thinking to the LLM is often a gravitational pull for following \inlinequote{the path of least resistance} [P4], which bypasses the cognitively demanding work of coming up with original solutions for creative problem solving. Prior research reveals various risks associated with such cognitive offloading, affecting both individual and group creativity~\cite{Zhou2025WhoET}. Ceding autonomy to AI could lead to a decline in internal cognitive abilities, such as critical thinking skills and memory retention~\cite{chakrabarty2024art}. More importantly, LLMs' bias toward statistically common responses constrains creative exploration, nudging users toward modal solutions that eventually converge users' ideas into similar repetitive solutions~\cite{girotra2023ideas}. Recently, Doshi et al.~\cite{doshi2024generative} found that using LLM assistants may have benefits for individual creativity, but it reduces diversity when looking at groups of individuals. This is further echoed by our participant [P22], who exclaims that with the large-scale adoption of LLMs leading creative effort, we will be left with \inlinequote{a population of humans less creative!}

To design effective creativity support tools that leverage generative AI capabilities, we must understand how LLMs impact the entire creative process of users. Existing creativity support tools mostly augment the ideation stage of creative thinking by suggesting and expanding on users' ideas~\cite{reicherts2025ai, DiFede2022The, Suh2023LuminateSG}. Others explored supporting idea generation through planning and reviewing tasks that require creative thinking~\cite{Chakrabarty2024Creativity}. A recent work by Suh et al.~\cite{Suh2023LuminateSG} aims to support design space exploration through structured frameworks. However, these tools mostly overlook where users need support in other stages of creative problem solving, like verifying and refining ideas and precisely implementing them~\cite{Wallas1970TheAO}. Designing support without considering where it is needed is inefficient and compounds the risk of cognitive decline and homogeneous solutions~\cite{Anderson2024Homogenization} by providing all users with a plethora of similar ideas.
In this paper, we study the impact of using LLMs on creative thinking in problem-solving using a counterbalanced within-subject user study with N=20 participants. We situated our study in the domain of programming, and each participant completed four programming tasks during the study. This is an ideal context for several reasons. 
First, creativity is fundamental to programming practice and critical for software 
engineering outcomes. Creative problem-solving enables programmers to devise novel 
algorithms, architect scalable systems, and develop innovative solutions to complex technical challenges~\cite{Amini2024Coding} and it is one of the key qualities of software developers~\cite{li2020distinguishes,li2015makes} as they need to think flexibly to diagnose unexpected behaviors and reimagine code structures. As software systems grow in complexity, the ability to creatively synthesize existing components, adapt solutions across domains, and generate original approaches becomes increasingly valuable for both individual productivity and organizational innovation.

Second, programming enables us to study the impact on creativity from both process and product perspectives. In some cases, programmers are driven by the broader objective of developing innovative (creative) products. In others, there is value in expressing creativity during the program development process, where users can synthesize new solutions by creatively combining traditional methods. While existing investigations on the impact of LLM on creativity in art or writing focus solely on evaluating the product~\cite{doshi2024generative, Zhou2025WhoET,Wan2023ItFL}, our programming-based study enables us to examine both the process and product of creativity. Programming also offers more measurable outcomes compared to more subjective fields, such as art or writing. Finally, programming holds high ecological validity as LLM-based coding assistants are widely adopted and used by millions of developers in real life~\cite{ArsTechnica2025DeveloperSurvey}. Studying programming, therefore, provides timely and representative insights into a real-world, high-stakes instance of human-AI creative collaboration. As such, it serves as a suitable testbed for understanding how LLMs impact the entire creative process.

To investigate the effect of LLM use on the creative process, we observed participants as they performed four programming tasks under two randomly assigned conditions: \textit{LLM-Assisted} and \textit{Unassisted}. We seek to answer the following research questions:
\begin{itemize}
\item RQ1: How does LLM assistance affect the creativity of the solutions?
    \item RQ2: How does LLM usage impact creativity during programming?
    \item RQ3: How does LLM support creativity in LLM-assisted sessions?
\end{itemize}
 
\rrevised{Qualitative and quantitative analyses revealed that participants engaged in idea generation significantly more often and for longer in the unassisted condition ($p_{holm} < 0.001$), and participants described working without assistance as prompting deeper engagement with the problem.}

\rrevised{Beyond the creative process, LLM use also shaped the solutions themselves. LLM-assisted code was functionally and syntactically better and contained a similar number of unique ideas, but for algorithmic tasks, LLM-assisted solutions were significantly more similar to one another, indicating a narrower solution space (RQ1).}
\new{ Our findings suggest that while using LLMs provided productivity gains, participants did not utilize the gained time to engage in creative thinking.} In fact, participants experienced a significantly higher number of creative moments across all stages of problem-solving in the unassisted conditions (RQ2) compared to the LLM-assisted ones. However, this does not imply that LLM use is always harmful for creativity. Looking deeper into how participants used LLMs revealed four common collaboration modes that support different problem-solving styles and strategies: the \CMone, \CMtwo, \CMthree, and \CMfour collaboration models (RQ3). \new{Collaboration modes where participants preserved idea generation agency (LLMs used as implementer and verifier) experienced more creative moments and generated diverse solutions compared to modes where they ceded thinking agency.}

Based on our findings, we discuss direct design implications for developing creativity support tools for programming. We discuss how to balance the tensions between cognitive effort and the creativity of solutions by providing dynamic and selective support that prevents users from falling into the path of least resistance. Additionally, we also discuss key design considerations for building LLMs that support creativity.

%% file: Sections/RW.tex
\section{\rrevised{Background and Related Work}}

\subsection{\rrevised{Creativity in Programming and the Shifting Role of Developers}}

\rrevised{Programming is often methodical, relying on established knowledge, techniques, and heuristics, but it also requires creativity~\cite{Amini2024Coding,Graziotin2013The,Groeneveld2021Exploring}. Developers distinguish routine tasks from creative work, viewing creativity as solving non-trivial, complex problems in new or personally new ways, and report that it is not uniformly needed across programming but concentrated in specific stages of problem solving such as brainstorming or problem decomposition~\cite{Groeneveld2021Exploring,Graziotin2013The,Groeneveld2023Students'}. Higher creativity is in turn associated with more diverse coding strategies and richer solution spaces~\cite{Amini2024Coding,Njoku2024Innovating,Inman2024Developer}.}

In contrast, later stages such as implementing well-established solutions, applying known algorithms, or following best practices tend to be less creative. Developers report low creativity when performing repetitive tasks, copying code, or executing tightly defined instructions \cite{Groeneveld2021Exploring,Graziotin2013The,Groeneveld2023Students'}, and systematic reviews find creativity emphasized more in requirements and design than in routine phases \cite{Lasisi2016Creativity,Graziotin2013The}.

In this work, we focus on exploring the productive creativity in programming, where creative thinking leads to novel ideas that drive more effective or efficient solutions.

With the rise of LLM-based coding assistants, the role of programmers is beginning to shift. Developers increasingly spend more time reviewing, evaluating, and adapting AI-generated code rather than writing code from scratch \cite{Bird2022Taking}. Emerging visions of software development similarly suggest a transition toward developers acting as orchestrators of AI-driven workflows, focusing on higher-level decision making and design \cite{Qiu2024From}.
\rrevised{As a result, responsibility for ensuring novelty, diversity, and context-appropriate solutions remains with the developer~\cite{Huang2023Bias,Wong2023Natural}, shifting the locus of creativity away from manual code writing toward problem framing, exploration of alternatives, and critical evaluation of generated solutions. Rather than replacing creativity, AI tools may amplify the need for it, making creative thinking a key component of productivity in modern programming.}

\subsection{Understanding Creativity}
\label{sec:frameworks}
\emph{Creativity frameworks.} Creativity has been conceptualized through a range of frameworks, each emphasizing different aspects of this concept. Recent works have investigated creativity through the creativity process~\cite{Marron2019Measuring,Pinkow2022Creative}. The Wallas Stage Model~\cite{Wallas1970TheAO} presents creativity as a sequential, cyclic process: individuals first engage in Preparation by gathering knowledge, then allow ideas to incubate subconsciously, experience sudden insight during Illumination, and finally refine and validate their ideas in the Verification stage. This model highlights a temporal progression from initial confusion to eventual understanding. Beyond the sole creative process, some frameworks have been proposed, such as the famous 8P Framework~\cite{Sternberg2021An}. This framework includes a broader perspective into the creativity study, situating creativity within eight interrelated dimensions: Purpose, Press (environmental influences), Person, Problem, Process, Product, Propulsion (motivation), and Public. This approach illustrates how individual traits, contextual factors, and social dynamics collectively shape creative outcomes.

Similarly, in a more recent work, the Dynamic Creativity Framework~\cite{Corazza2022The} conceptualizes creativity as an ongoing, context-sensitive process. It highlights continuous interactions among motivational drive, information acquisition, idea generation, and evaluation, while acknowledging the influence of cognitive, emotional, and environmental factors on creative activity. Boden’s Hierarchy~\cite{Wiggins2006A} emphasizes the types of creativity rather than the process itself, distinguishing among combinational creativity (recombining existing ideas), exploratory creativity (expanding the boundaries of established rules), and transformational creativity (generating entirely novel rules or conceptual spaces). This framework is applicable to both human and artificial systems.

\rrevised{Although creativity has been studied across various domains and from multiple perspectives, such as process, type, and product, these frameworks offer limited guidance for understanding how individual creativity is affected, in both process and product, in the context of LLM-assisted problem solving.}

\subsection{LLM-based Creativity Support Tools in Human-AI Interactions}

\emph{LLMs applications in Creativity Support tools.} Large language models (LLMs) are increasingly employed in creativity support tools (CSTs) to facilitate ideation, brainstorming, content generation, and design exploration~\cite{10.1145/3586183.3606719}. Platforms such as Luminate~\cite{Suh2023LuminateSG} and Supermind Ideator provide structured design spaces and scaffold~\cite{10.1145/3613904.3642335} the creative problem-solving process, enabling users to generate, evaluate, and synthesize ideas more effectively ~\cite{Radensky2024ScideatorHS,heyman2024supermind, shaer2024ai,DiFede2022The}.

Beyond individual use, LLMs are integrated into collaborative environments (e.g., Collaborative Canvas, Jamplate) to support group ideation, facilitate knowledge exploration, and enhance shared creative workflows~\cite{Dunnell2023LatentLL,Gonzalez2024CollaborativeCA}. \new{Although recent research on collaborative creative ideation with LLMs shows that users’ misunderstandings of model capabilities can reduce creative idea generation~\cite{10.1145/3706598.3713886}, this finding reinforces our motivation to better understand the creative dynamics of programmer–LLM interactions}. In domains such as programming and design, LLMs assist users in exploring alternative solutions, automating repetitive scripting tasks, and lowering barriers to creative coding ~\cite{zamfirescu2025beyond,rietschel2024mediating,wang2024exploring}).

\emph{Impact of LLMs on Creativity}. 

Empirical research indicates that LLMs can increase both the quantity and richness of ideas generated, particularly benefiting less experienced users, and can enhance the perceived creativity of outputs ~\cite{Anderson2024Homogenization,heyman2024supermind,doshi2024generative,lim2024rapid} However, concerns about homogenization remain: LLMs may inadvertently encourage convergence on similar solutions, reducing collective originality and diversity ~\cite{Anderson2024Homogenization,doshi2024generative,mohammadi2024creativity}. Importantly, the impact of LLMs on creative outcomes depends on interaction modalities—for example, whether the LLM functions as a “ghostwriter” or a sounding board—and on the presence of scaffolds or structured interaction~\cite{heyman2024supermind,chen2024large,wang2024exploring}.

\rrevised{Across this literature, LLM-based CSTs primarily target, and are primarily evaluated on, the idea generation stage; the remaining stages of the creative process have received considerably less attention. This motivates our study: by examining the impact of LLMs at each stage of creative problem solving, we identify where LLMs currently support human creativity effectively and where they fall short.}

%% file: Sections/Method.tex
\section{Methodology}

\subsection{Study Design}

\begin{figure}
\centering
\includegraphics[width=\linewidth]{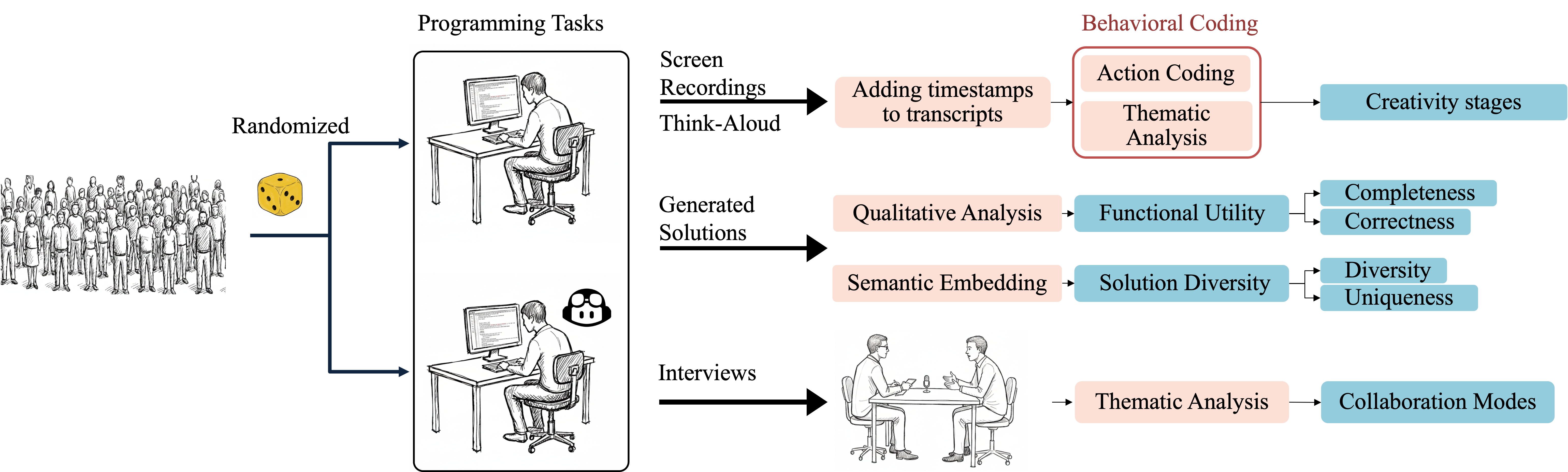}
\caption{Experimental Design and Analysis. An overview of the randomized study comparing Unassisted versus LLM-Assisted creative programming. The pipeline details data collection methods (screen recordings, Python code snippets, interviews) and the subsequent mixed-methods analysis of programming stages, code quality metrics, and collaboration modes.}
\Description{Flow diagram of the study design and analysis. On the left, a large group of people is randomly assigned (shown by a die) into two conditions, each depicted as a participant coding at a computer: an Unassisted participant (top) and an LLM-assisted participant with an AI icon (bottom). On the right, the collected data feed four analysis streams, with pink boxes marking methods and blue boxes marking the resulting constructs. First, adding timestamps to transcripts feeds Behavioral Coding (which combines Action Coding and Thematic Analysis) to produce Creativity stages. Second, Qualitative Analysis feeds Functional Utility, yielding Completeness and Correctness. Third, Semantic Embedding feeds Solution Diversity, yielding Diversity and Uniqueness. Fourth, a retrospective interview (two people at a table) feeds Thematic Analysis to produce Collaboration Modes.}
\label{fig:diagram}
\end{figure}

In this paper, we conducted \revised{an observation-based within-subject experiment with 20 participants,} followed by retrospective interviews. 

An overview of our crossed-design experimental design is shown in Figure~\ref{fig:diagram}.
Each experiment comprised two sessions, one per condition. In one session, participants completed two programming tasks under the \textit{Unassisted condition}; in the other, they performed different programming tasks in the \textit{LLM-assisted} condition. To isolate the effect of LLM access, participants in both conditions were restricted to using only the provided IDE. During the \textit{Unassisted} condition, participants were not allowed to use any external resources; all coding activity took place within Visual Studio Code. We intentionally imposed this restriction for the control group to isolate the specific impact of LLM coding assistance (GitHub Copilot), avoiding confounds like web search or online documents. During the \textit{LLM-assisted} condition, participants had access to GitHub Copilot \revised{with GPT 4.1 model} via its installed IDE extension but were not allowed to use any other external resources. \revised{To make sure the type of model in the LLM does not cause a confounding factor all participation only had access to the same model.} \rrevised{Copilot was left at its default Visual Studio Code configuration except that inline autocompletion was disabled (Section~\ref{sec:method:pilots}), so that every LLM interaction was explicitly initiated by the participant through Copilot Chat, and each participant started from a new, empty chat with no history from prior participants. The full environment and Copilot configuration are reported in the supplementary material.}

To control for order effects, we counterbalanced the condition sequence: half of the participants completed the Unassisted condition first and then the LLM-assisted condition, while the others completed them in the opposite order.
All sessions were conducted in person in a private room to help participants feel comfortable and reduce awareness of being observed. We employed an observational study protocol to capture direct, real-time insights into participants' creative programming while minimizing self-report bias.

\begin{figure}
\centering
\includegraphics[width=\linewidth]{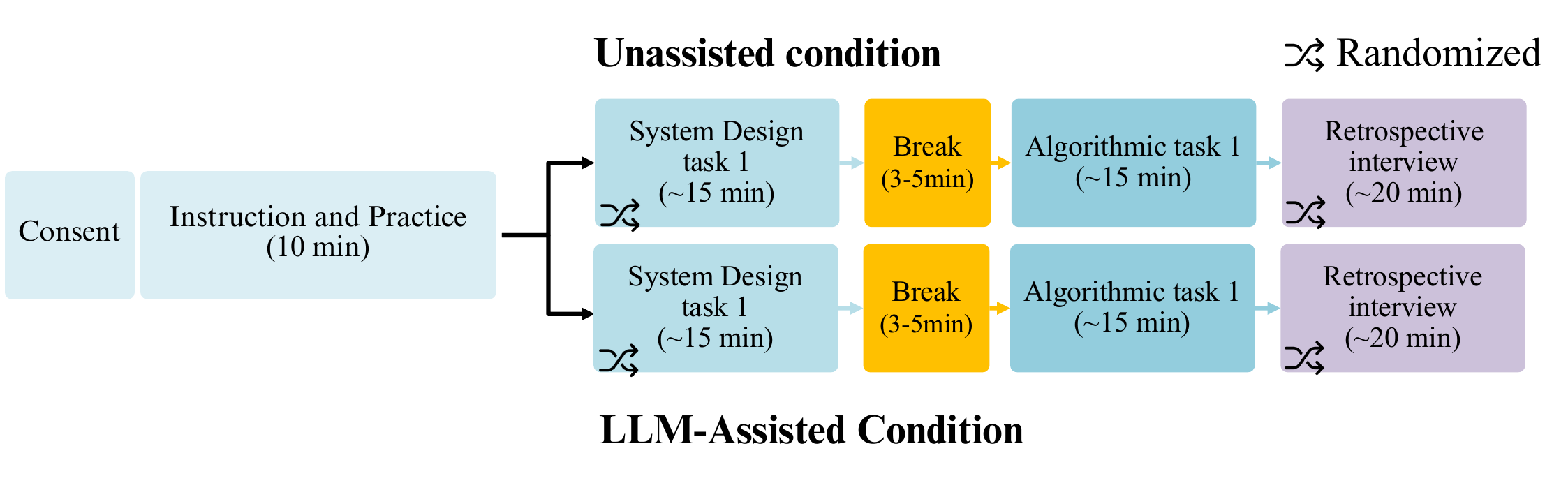}
\caption{\revised{Participant Study Protocol.} After recruitment, participants signed consent forms and familiarized themselves with the coding environment through a practice session. The order of conditions and task types was randomized.}
\Description{Left-to-right flowchart of the session protocol. It begins with Consent, then Instruction and Practice (10 minutes), after which the flow splits into two parallel randomized tracks: the Unassisted condition (top) and the LLM-Assisted condition (bottom). Each track runs the same sequence of four steps: a System Design task (about 15 minutes), a Break (3 to 5 minutes), an Algorithmic task (about 15 minutes), and a Retrospective interview (about 20 minutes). Shuffle icons on the task and interview boxes, together with a "Randomized" legend, indicate that the order of conditions and task types was randomized across participants.}

\label{fig:Participant_protocol}
\end{figure}

\subsection{Participants}

We recruited 23 participants through snowball and convenience sampling from undergraduate and graduate students with varying academic levels. Our inclusion criteria required participants to have familiarity with LLM-based code generation tools and experience in Python programming. During data collection, one participant (P8) did not wish to continue with four assigned tasks, and data corruption affected two additional participants' recordings (P1, P2). We excluded these three participants entirely from our analysis to maintain balanced datasets across all remaining participants and preserve the integrity of our comparative analyses. The demographic information of our final 20 participants is shown in Table~\ref{tab:participants}.
\begin{table}[htbp]
\centering
\caption{Participant Demographics (N=20)}
\label{tab:participants}
\begin{tabular}{cccccc}
\toprule
\textbf{PID} & \textbf{Gender} & \textbf{Age} & \textbf{Prog. Exp.(yrs)} & \textbf{Education Level} & \textbf{LLM Usage} \\
\midrule
3 & Male & 31 & 12 & PhD & Often \\
4 & Male & 23 & 8 & MS & Sometimes \\
5 & Male & 23 & 7 & MS & Sometimes \\
6 & Male & 24 & 4 & MS & Sometimes \\
7 & Male & 37 & 7 & MS & Rarely \\
9 & Non-binary & 23 & 7 & PhD & Sometimes \\
10 & Female & 25 & 8 & PhD & Often \\
11 & Male & 25 & 3 & MS & Rarely\\
12 & Female & 24 & 8 & MS & Always\\
13 & Male & 22 & 6 & MS & Often \\
14 & Male & 29 & 8 & PhD & Sometimes \\
15 & Male & 29 & 5 & PhD & Always \\
16 & Female & 32 & 10 & PhD & Always \\
17 & Female & 25 & 10 & PhD & Often \\
18 & Female & 25 & 7 & PhD & Sometimes \\
19 & Female & 29 & 9 & PhD & Always \\
20 & Male & 21 & 4 & BC & Rarely \\
21 & Female & 30 & 5 & PhD & Always \\
22 & Male & 24 & 4 & MS & Sometimes \\
23 & Male & 26 & 10 & PhD & Always \\
\bottomrule
\end{tabular}
\end{table}

\subsection{Tasks}
 Each participant completed four programming tasks across two conditions. To control for order effects, task orders were randomized. All programming questions were implemented in Python. Two types of programming tasks were defined: \Atask were programming questions that included explicit space or time complexity requirements, while \Btask had no such complexity constraints and asked participants to design an application or feature within a given implementation context. We devised \revised{four} \Atask (A1, A2, A3, A4) and four \Btask (B1, B2, B3, B4). Each condition contained one Algorithmic task and one System Design task. Although the order of task presentation was randomized, we ensured that each participant performed one of the \Atask and one of the \Btask in each condition. \rrevised{In total, we collected 80 solutions from participants: 40 solutions to Algorithmic tasks and 40 for System Design tasks, with 20 of each task type in each condition.}
 \rrevised{Tasks within each type were designed to be comparable in scope and difficulty and were randomly assigned, so individual tasks were not evenly distributed across conditions; per-task counts are reported in the supplementary materials.}

\subsection{Procedure}
Detailed procedure is illustrated in Figure~\ref{fig:Participant_protocol}. Participants were instructed to think aloud while completing the tasks, and both their audio and screen activity were recorded. Each participant was given a baseline of 15 minutes to finish the task, but they were not rushed, and they could exceed 15 minutes as long as they wanted to complete the task. Each participant would start the task by seeing the task problem statement in a Python executable file (.py); they were asked to use the same file for implementing their solution. After completing a task, participants notified the researcher, and the researcher would pause all the recordings. 

\subsubsection{Self-reported creative moments.} Because the number of creative or "aha" moments is crucial in creativity research, we implemented a pop-up prompt every five minutes asking participants whether they had experienced any "aha" moments during the preceding interval. This self-reported measure of aha moments was later used by researchers to verify the creative moments they observed from the recordings.

\subsubsection{Retrospective Interviews.} At the end of each session, one researcher interviewed participants. These interviews explored: 1) problem-solving strategies that participants employed in each condition, 2) how LLM usage impacted their creativity in problem-solving, and 3) whether they experienced any aha or creative moments during the observation. We asked the following three core questions in the interview:
\begin{itemize}
\item How did you approach solving the problems?
\item Did the LLM offer strategies or approaches you wouldn't have thought of?
\item Did you have any "aha" moments? What triggered them?
\end{itemize}

\subsubsection{Pilots}
\label{sec:method:pilots} We observed in pilots that GitHub Copilot's automatic code completion feature created unintended interference with participants' creative processes. Since all coding tasks were short and conducted within single files, the autocompletion suggestions often provided complete or near-complete solutions before participants could engage in independent creative coding. Participants reported frustration with this feature, noting that the premature suggestions introduced unwanted cognitive bias and constrained their creative exploration. \revised{As a result, we disabled GitHub Copilot's default autocompletion feature for all participants in the main study.} \revised{Additionally, pilot participants noted that the wording of some tasks was ambiguous and made certain tasks disproportionately difficult; we revised the phrasing of those tasks accordingly prior to the main study. For example, one task's ordering condition was clarified from \textit{find three values that appear in the same relative order in both arrays''} to explicit positional notation (\textit{``pos1[x] < pos1[y] < pos1[z]''}).}

%% file: Sections/Analysis.tex
\section{Analysis}
In this section, we explain how we analyzed the collected data.

\subsection{Theoretical Framework}

\revised{We adopted Wallas's model as our theoretical framework because it remains the most widely used and well-established framework in creativity research~\cite{Corcuera2024The,SadlerSmith2015Wallas,Setiawani2019The,Yanti2018The,Maharani2017Creative}. Among process-oriented creativity frameworks, Wallas's four stages offer broad coverage of the creative process of creative problem solving. This structure also aligns naturally with the nature of programming as a problem-solving activity.}
\rrevised{As described in Section~\ref{sec:frameworks}, Wallas's model comprises four stages: Preparation, Incubation, Illumination, and Verification. Programming inherently unfolds as a structured problem-solving process~\cite{10.1145/2858036.2858252,Demir2025The}, so we adapted these stages to the programming context as follows:}

The \textit{Preparation} phase corresponds to requirement analysis when participants read and comprehend the problem specification. \textit{Idea Generation} represents the phase where participants evaluate different approaches, select appropriate data structures, and formulate solution strategies. We subdivided Wallas' Verification stage into \textit{Implementation} (translating algorithmic designs into executable code) and \textit{Verification} (debugging and validating that code) to reflect the distinct cognitive activities and iterative cycles observed in programming.
Contrary to Wallas' definition of Illumination as a stage, we observed \textit{Illumination} as discrete moments when participants experienced an ``aha'' moment or epiphany—realizing the algorithmic approach, identifying a logical bug, or recognizing computational inefficiency. These insights represent moments of problem-space restructuring, a cognitive process associated with learning and skill development. We acknowledge that self-reported ``aha'' moments are subjective and that the feeling of insight does not necessarily imply correctness or novelty~\cite{Salvi2016InsightSA}. To mitigate this, we triangulated participants' reports with multidimensional markers: the presence of an \textit{impasse} followed by \textit{suddenness} and \textit{positive affect}~\cite{Topolinski2010,Sandkühler2008Deconstructing,Stuyck2021The}, distinguishing genuine cognitive restructuring from simple memory retrieval. Throughout this paper, we refer to these adapted stages as programming stages, emphasizing their grounding in programming-specific creative activity.

\subsection{Behavioral Coding}

\label{sec:analysis:coding}

\begin{figure}
    \centering
    \includegraphics[width=\linewidth]{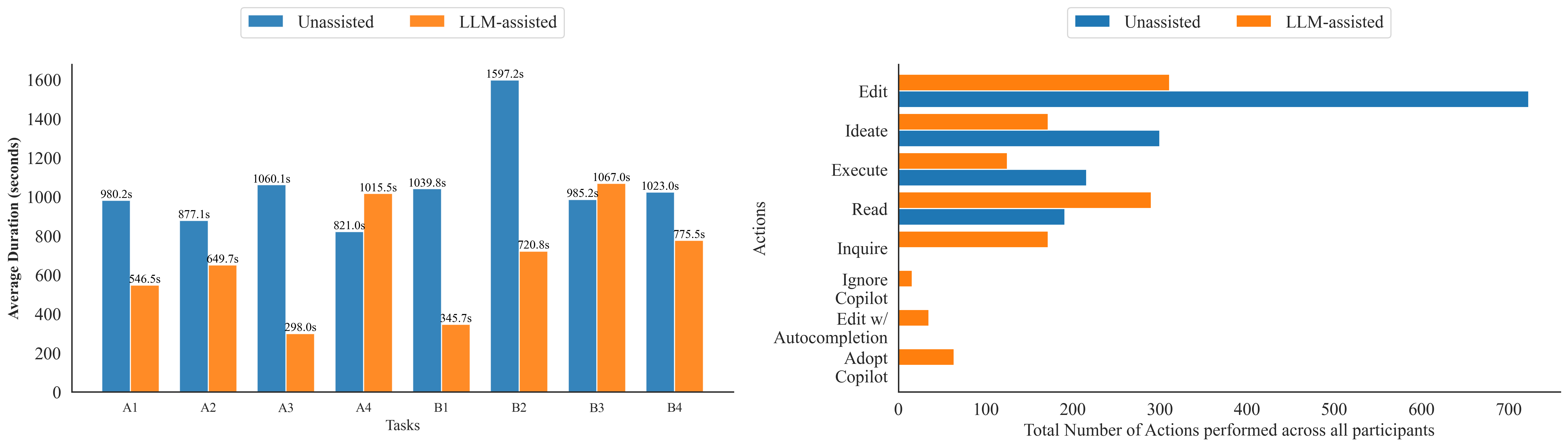}
    \caption{Comparison of all participants (a) total time spent on each task across conditions (b) total frequency of each actions }
    \label{fig:overview}
    \Description{Two bar charts comparing conditions, with blue for Unassisted and orange for LLM-assisted. Panel (a) is a grouped vertical bar chart of average task duration in seconds for the eight tasks A1 to A4 and B1 to B4; unassisted bars are generally taller, indicating longer durations. Panel (b) is a horizontal bar chart of the total number of each action type (Edit, Ideate, Execute, Read, Inquire, IgnoreCopilot, Edit with autocompletion, AdoptCopilot) aggregated across participants; Edit, Ideate, Execute, and Read dominate, and the Copilot-specific actions appear only in the LLM-assisted condition.}
\end{figure}

To identify the stages mentioned above in our observational data, we developed a systematic coding scheme. Following Chattopadhyay et al.~\cite{chattopadhyay2020tale}, we categorized participants' interactions into eight programming actions: \textit{Read}, \textit{Edit}, \textit{Execute}, \textit{Ideate}, \textit{Inquire}, \textit{AdoptCopilot}, \textit{IgnoreCopilot}, and \textit{Edit with Autocompletion}. An action is defined as a discrete step taken by a participant to pursue a specific, consistent goal. Detailed definitions for each action are in Table~\ref{tab:action_coding}. \rrevised{In total, we coded 2,591 actions across the two conditions (an average of 130 per participant): \revised{$N_u = 1416$} in the \emph{Unassisted} condition, where participants had no access to external resources or LLM assistants, and \revised{$N_l = 1175$} in the \emph{LLM-assisted} condition, where participants had access to an LLM. On average, each participant spent 63.49 minutes across all sessions, \revised{with completion times ranging from 36.18 to 94.26 minutes}, amounting to $T_c = 703.4$ minutes in the Unassisted condition and $T_t = 566.3$ minutes in the LLM-assisted condition (Figure~\ref{fig:overview}(\modified{a})). The most common actions were editing code or prompts ($n = 1034$), reading code, output, or LLM responses ($n = 481$), ideating about solutions ($n = 472$), executing code ($n = 341$), and prompting the LLM ($n = 172$). Figure~\ref{fig:overview}(\modified{b}) shows the frequency of each action type across the two conditions, \modified{with the horizontal axis representing the aggregated number of actions across all participants.}}

\begin{table}[htbp]
\centering
\caption{Action Coding Definitions}
\label{tab:action_coding}
\begin{tabularx}{\textwidth}{>{\bfseries}l X}
\toprule
\textbf{Action} & \textbf{Definition} \\
\midrule
Read & Examining information from artifacts (e.g., code, documentation, terminal output). \\
\addlinespace
Ideating & Constructing mental models of plans or engaging in idea generation activities for developing problem solutions, or identifying issues and bugs in strategies. \\
\addlinespace
Edit & Any changes made directly to the code without help from Copilot. \\
\addlinespace
Edit with autocompletion & Any change made directly to code or artifacts with help from Copilot, including copy pasting the code from Copilot chat, and modifying the suggestion. \\
\addlinespace
Execute & Compiling, and/or running code. \\
\addlinespace
Inquire & Prompt Copilot. \\
\addlinespace
AdoptCopilot & Decide to follow instructions and/or accept the suggestion from Copilot. \\
\addlinespace
IgnoreCopilot & Disregard instructions or codes generated by Copilot. \\
\bottomrule
\end{tabularx}
\end{table}
Two researchers performed behavioral coding at each timestamp, relying on (1) participants' verbalization, (2) actions being performed, and (3) observed problem-solving behavior. Using an inductive, open-coding approach, researchers coded each participant through negotiated agreement, iteratively updating the codebook with new codes and definitions until saturation was reached after 14 participants. This process yielded 66 unique codes. Researchers then reviewed and merged codes with similar meanings (e.g., \textit{brainstorming} and \textit{exploring alternative solutions}), consolidating to 51 final unique codes. The complete codebook is available in the supplementary materials\revised{\footnote{https://figshare.com/s/f39eed0e7e30d6ae6a32}}.
Each code was mapped to the corresponding creativity stage defined in our framework. For example, "understanding the problem statement" mapped to Preparation, while \textit{``finding possible solution(s) with thinking''} mapped to Idea Generation. The code names indicated whether participants used an LLM to perform the specific behavior. Through this mapping, we assigned 11 codes to Preparation, 13 to Idea Generation, 3 to Illumination, 7 to Implementation and 17 to Verification. We used these mapped stages to identify the frequency and duration of each creativity stage using the timestamps. 

\rrevised{To validate open-coding consistency, two researchers independently
coded a randomly selected 60-minute subset of the data in each of three rounds,
until Cohen's $\kappa$ exceeded our preset threshold of 0.85. These 51 codes were mapped directly onto the framework's programming stages; no thematic grouping was applied at this step.}

\rrevised{For our analysis across conditions, we use Aligned Rank Transform (ART) ANOVA~\cite{Wobbrock2011TheAR}. Because our study is a within-subjects crossed-design, where each participant completes tasks across all conditions and task types, observations within a condition or task type are not independent. We therefore modeled participants as a random intercept, with condition and task type (\Atask{} or \Btask{}) as fixed effects. We further ensure that our collected data meet all the required assumptions for ART ANOVA (including non-normality, tested using the Shapiro-Wilk test; $p<0.05$). ART ANOVA's rank-based transformation makes no normality assumption while still supporting factorial effects, interactions, and random effects. As multiple outcomes were tested within each family, we further applied the Holm--Bonferroni step-down procedure~\cite{Holm1979ASS} to control the family-wise error rate without assuming independence between tests.}

\subsubsection{Creative Moment Detection.} 
\revised{As a part of the creativity process, we identified the creative moments that participants experienced during the experiment. }
\rrevised{Creative moments have been widely used as an indicator of creative presence in controlled experiments~\cite{Sprugnoli2017Neural,Carpenter2019The,Laukkonen2018How}; in our study, they serve as one of several indicators, alongside stage-level behavioral coding and solution diversity.}
To identify creative moments, we coded every instance where participants expressed an "aha" or similar reaction in their think-aloud verbalization. We implemented a pop-up prompt at five-minute intervals, asking participants whether they had experienced any aha moments (explained during pre-experiment warm-up). If participants responded "yes," we examined the corresponding interval to verify that the moment had been recorded. Conversely, if participants reported no aha moments but the researcher had identified one, this was addressed during retrospective interviews to confirm which moments actually occurred. \revised{Only moments confirmed by both the researcher and the participant were retained; researcher-only annotations were discarded to preserve validity}. This multi-source triangulation, combining verbal exclamations, self-reported pop-up responses, observable behavioral shifts, and post-interview verification, ensured we captured genuine insight moments rather than simple information recall. \rrevised{Candidate moments were double-coded under the same protocol (Cohen's
$\kappa > 0.85$) before participant confirmation.} \rrevised{Because creative-moment counts are non-normal and each participant contributed repeated observations, we compared them across conditions using ART ANOVA with participant as a random intercept, applying Holm--Bonferroni correction across outcomes (see Section~\ref{sec:analysis:coding}).}

\subsubsection{Analyzing Copilot Usage patterns}

To examine how participants' collaboration patterns with LLMs influenced their creativity, two researchers conducted an inductive open-coding analysis of retrospective interview transcripts using iterative, negotiated agreement. We coded instances where participants elaborated on their problem-solving strategies in the Unassisted condition and described how the LLM affected those strategies in the LLM-assisted condition. \rrevised{Open coding of the transcripts yielded 21 unique codes, with Cohen's $\kappa > 0.85$. Two researchers then applied thematic
analysis~\cite{braun2006using} to these 21 codes, resolving all disagreements
through negotiated agreement, and arrived at four themes, which we report as
collaboration modes: \CMone, \CMtwo, \CMthree, and \CMfour.}

\subsection{Code Artifact Analysis}

We conducted an initial analysis to identify differences in code features across two conditions. \revised{We used the} Python frameworks \textit{Radon} and \textit{AST} \revised {to extract} eight metrics across syntactic, structural, and stylistic dimensions: number of function calls, number of external libraries used, lines of code, logical complexity, code maintainability, use of comprehensions, recursive functions used, and average function length. These metrics are \revised{language-agnostic and} standard in the literature for code comparison and LLM-generated code quality evaluation~\cite{Dou2024What's,Zheng2024Beyond}. \revised{We computed them using Python-specific tooling because our generated solutions were in Python.}
Following the typical framework in creativity research~\cite{amabile1983social,woodman1993toward,harvey2023toward}, we analyzed code solutions along two dimensions: functional utility and solution diversity.

\subsubsection{Functional Utility}
Participants' solutions varied widely in form. While LLM-assisted solutions were almost always complete, unassisted solutions sometimes included pseudocode, natural language descriptions, or code with syntax errors.

Two expert raters independently assessed each of the 80 solutions along two dimensions:
\emph{Correctness} captures whether the underlying algorithmic idea is valid, regardless of syntactic correctness or executability. Solutions were rated as \textit{Correct} (core logic would produce the right output given proper implementation), \textit{Partially Correct} (sound approach but flawed edge case or secondary logic handling), or \textit{Incorrect} (flawed or missing fundamental approach).
\textit{Fidelity} captures implementation completeness independent of correctness: (1)~\textit{Pseudocode or Natural Language} (structured description without Python syntax), (2)~\textit{Incomplete Implementation} (structural scaffolding with gaps, placeholders, or missing components), (3)~\textit{Near-Complete Code} (largely finished but with syntax or runtime errors), and (4)~\textit{Executable Code} (runs without errors).
Raters first coded all solutions independently, then resolved disagreements through discussion, reaching high inter-rater agreement (Cohen's $\kappa > 0.85$).


\subsubsection{Solution Diversity}
We measured diversity across code artifacts using semantic similarity and main idea novelty analysis.

\emph{Semantic Similarity Analysis.} We adopted a semantic embedding approach using the Voyage Code 3 embedding model, which projects different surface representations (executable code, pseudocode, natural language) into a shared vector space where proximity reflects semantic similarity in the algorithmic approach. This addresses the limitation of traditional syntactic metrics like CodeBLEU~\cite{ren2020codebleu}, which assume well-formed, parseable code and would fail on pseudocode or natural language descriptions present in our unassisted solutions.

For each of the 80 solutions, we generated a 1024-dimensional embedding vector and computed pairwise cosine similarity between all solution pairs \rrevised{to the same question} within each condition, separately for \Atask{} and \Btask{}. Higher pairwise similarity indicates solutions are more alike (narrower solution space); lower similarity indicates greater diversity (broader exploration). \rrevised{Restricting comparisons to the same question is necessary because solutions to different questions are dissimilar regardless of condition, and the number of Unassisted and LLM-assisted solutions differs across questions.}

We restricted the analysis to solutions rated as Correct or Partially Correct. This filtering ensures observed diversity differences reflect genuine variation in valid problem-solving strategies rather than noise from erroneous attempts---incorrect solutions would naturally differ from correct ones, artificially inflating diversity. \rrevised{Because each participant contributed solutions to both conditions, and because each solution enters many pairwise comparisons, the similarity values are not independent. We therefore summarized each solution by its mean cosine distance to the other solutions to the same question in the same condition---averaging these values within a condition reproduces its mean pairwise distance exactly---and compared conditions with a linear mixed-effects model on the log scale, with crossed random intercepts for participant and question. Significance was assessed by permuting condition labels within each question and refitting the model (1,000 permutations), which assumes only that condition labels are exchangeable within a question and that participants are independent.}

\emph{Idea Novelty Assessment.} Two researchers qualitatively analyzed the primary strategy participants attempted to implement. Through negotiated agreement, they extracted the main idea of each solution, then rated how original each approach was within each task relative to others who completed the same task. These originality ratings were compared across conditions. In creativity literature, originality is widely viewed as a core indicator of creative performance~\cite{Mayseless2015Generating,Runco1993Judgments,Acar2017Ingredients}.

%% file: Sections/RQ1.tex
\section{RQ1: How does LLM assistance affect the quality and diversity of solutions?}
\label{sec:rq1}

\emph{\rrevised{Code Characteristics.}} While LLMs can generate more functional code, how does this code compare in structure, resource use (e.g., memory), and quality? To compare the code characteristics, we employed a Python-based automated framework that extracts standard code features used to measure code quality in software engineering. We analyzed eight key metrics across syntactic, structural, and stylistic dimensions (Table~\ref{code_metrics_scores}).
For comparing \textit{Syntactic and Structural Features}, we measured functional complexity through the number of function calls and average function length, logical complexity via branching patterns and independent code paths, and Python-specific sophistication through the use of comprehensions and recursive functions. To compare \textit{External Resource Usage}, we tracked the number of imported libraries as an indicator of leveraging external tools for novel solutions. Finally, to compare \textit{Code Quality}, we assessed lines of code (as a measure of conciseness versus verbosity), comment density, and a composite maintainability score reflecting readability and structural clarity. \rrevised{As the eight code metrics are non-normal and drawn from repeated observations per participant, we compared them across conditions using ART ANOVA and Holm--Bonferroni correction.}

Our analysis using ART ANOVA with Holm-Bonferroni correction reveals that LLM-assisted solutions were significantly more verbose (Lines of code: \rrevised{75.2 vs. 45.6}, \rrevised{$F(1, 47) = 10.41$, $p_{\mathrm{holm}} = 0.011$, $\eta_p^2 = 0.18$}) and included more function calls (\rrevised{16.76 vs. 8.00}, \rrevised{$F(1, 47) = 18.24$, $p_{\mathrm{holm}} < 0.001$, $\eta_p^2 = 0.28$}), suggesting longer and more fragmented or modular solutions. These solutions also demonstrated greater functional complexity (\rrevised{15.11 vs. 6.61}, \rrevised{$F(1, 47) = 11.45$, $p_{\mathrm{holm}} = 0.009$, $\eta_p^2 = 0.20$}). \rrevised{LLM-assisted code also scored lower on the maintainability index (76.1 vs 86.0, $F(1, 47) = 17.82$, $p_{\mathrm{holm}} < 0.001$, $\eta_p^2 = 0.28$).}
\begin{table}[h]
\centering
\caption{\rrevised{ART ANOVA Comparisons of Code Metrics Between Unassisted and LLM-assisted Conditions}}
\setlength{\tabcolsep}{4pt}
\begin{tabular}{lrrrrr}
\hline
\textbf{Metric} & \textbf{Unassist.} & \textbf{LLM-asst.} & \rrevised{$\boldsymbol{F(1, 47)}$} & \rrevised{$\boldsymbol{p_{\mathrm{holm}}}$} & \rrevised{$\boldsymbol{\eta_p^2}$} \\
\hline
Function Calls & 8.000 & 16.757 & \rrevised{18.24} & \rrevised{\textbf{**0.00075}} & \rrevised{0.28} \\
External Libraries Used & 0.061 & 0.216 & \rrevised{4.14} & \rrevised{0.14297} & \rrevised{0.08} \\
Lines of Code & 45.636 & 75.243 & \rrevised{10.41} & \rrevised{\textbf{*0.01143}} & \rrevised{0.18} \\
Logical Complexity & 6.606 & 15.108 & \rrevised{11.45} & \rrevised{\textbf{**0.00871}} & \rrevised{0.20} \\
Code Maintainability & 85.958 & 76.116 & \rrevised{17.82} & \rrevised{\textbf{**0.00077}} & \rrevised{0.28} \\
Comprehensions (List/Dict/Set) & 0.121 & 0.730 & \rrevised{6.15} & \rrevised{0.06721} & \rrevised{0.12} \\
Recursive Functions & 1.000 & 1.270 & \rrevised{0.00} & \rrevised{0.96458} & \rrevised{0.00} \\
Avg. Function Length (lines) & 8.601 & 11.441 & \rrevised{4.01} & \rrevised{0.14297} & \rrevised{0.08} \\
Non-executable Files (of 40) & 6 & 0 & -- & -- & -- \\
\hline
\end{tabular}
\label{code_metrics_scores}
\end{table}

\modified{These findings suggest that LLM-assisted solutions can indeed enhance productivity, enabling participants to write functionally better code by leveraging the sophisticated generative capabilities of the language models. This is especially valuable when writing solutions under time constraints. \rrevised{At the same time, these solutions scored lower on a static maintainability index and were structurally more complex. We measured maintainability as a static property of a single, short, self-contained artifact and did not observe these artifacts being extended or maintained, so whether these structural differences translate into downstream maintenance cost remains an open question.}}

\subsection{Impact of LLMs on Solution Correctness and Fidelity}
\label{sec:rq1:correctness}

\begin{table}
\centering
\caption{Distribution of correctness ratings by condition. Correctness reflects whether the underlying algorithmic idea is valid, regardless of whether the code is executable. A chi-square test confirmed a significant association between condition and correctness ($\chi^2 = 13.14$, \rrevised{$p < 0.05$}).}
\label{tab:correctness}
\begin{tabular}{lccc}
\toprule
 & \textbf{Correct} & \textbf{Partially Correct} & \textbf{Incorrect} \\
\midrule
Unassisted    & 18 & 11 & 11 \\
LLM-assisted  & 32 &  1 &  7 \\
\bottomrule
\end{tabular}
\end{table}

\rrevised{To compare solutions’ correctness and implementation fidelity across conditions, we used a chi-square test of independence, which tests whether the distribution of ratings is associated with condition without assuming normality.} Table~\ref{tab:correctness} reports the distribution of correctness ratings across conditions. Solutions generated with LLM assistance were significantly more correct ($\chi^2 = 13.14$, \rrevised{$p < 0.05$}), with 32 out of 40 rated as fully correct, compared to 18 out of 40 in the unassisted condition. Conversely, the unassisted condition produced a notably higher number of partially correct solutions (11 vs.\ 1) and incorrect solutions (11 vs.\ 7). 

\revised{In the unassisted condition, partial correctness largely reflected implementation gaps rather than conceptual misunderstanding; participants grasped the problem but lacked the syntactic or library knowledge to fully execute their approach.}
Importantly, the majority of unassisted solutions (29 out of 40) were still rated as correct or partially correct, meaning that participants' core problem-solving ideas were largely valid even when working independently.

\begin{table}[t]
\centering
\caption{Distribution of implementation fidelity ratings by condition. Fidelity captures how completely a participant's idea was translated into working code, independent of whether the underlying approach is correct. A chi-square test confirmed that LLM-assisted submissions were significantly ($\chi^2 = 17.06$, $p < 0.001$) more executable.}
\label{tab:fidelity}
\begin{tabular}{lcccc}
\toprule
 & \textbf{Pseudocode /} & \textbf{Incomplete} & \textbf{Near-Complete} & \textbf{Executable} \\
 & \textbf{Nat.\ Language} & \textbf{Implementation} & \textbf{Code} & \textbf{Code} \\
\midrule
Unassisted    & 7 & 7 & 7 & 19 \\
LLM-assisted  & 0 & 4 & 1 & 35 \\
\bottomrule
\end{tabular}
\end{table}

The fidelity dimension, reported in Table~\ref{tab:fidelity}, revealed a complementary pattern ($\chi^2 = 17.06$, $p < 0.001$). In the unassisted condition, only 19 out of 40 submissions were fully executable, with the remaining 21 distributed evenly across pseudocode or natural language descriptions (7), incomplete implementations (7), and near-complete code with minor errors (7). In contrast, LLM-assisted submissions were overwhelmingly executable: 35 out of 40 produced valid, runnable code, with only 4 incomplete implementations and 1 near-complete submission. No LLM-assisted participant resorted to pseudocode or natural language. This gap reflects a well-documented benefit of LLM code generation: the model's ability to translate high-level intent into syntactically valid, runnable programs. Participants who submitted non-executable code in the unassisted condition often expressed correct algorithmic ideas but lacked the recall of specific Python syntax or library functions needed to produce working code within the time limit.

Does this improvement come at the cost of solution diversity? Are LLM-assisted participants converging on the same correct approach, or are they exploring the solution space as broadly as unassisted participants? The distinction between having a correct idea and producing a fully functional implementation becomes central to our diversity analysis below, where we restrict the comparison to solutions with valid underlying approaches.

\subsection{Impact of LLMs on Solution Diversity}
\label{sec:rq1:diversity}

\subsubsection{Measuring Diversity with Semantic Code Embeddings.}
\label{sec:rq1:pairwise}

\begin{figure*}[t]
\centering
\includegraphics[width=\textwidth]{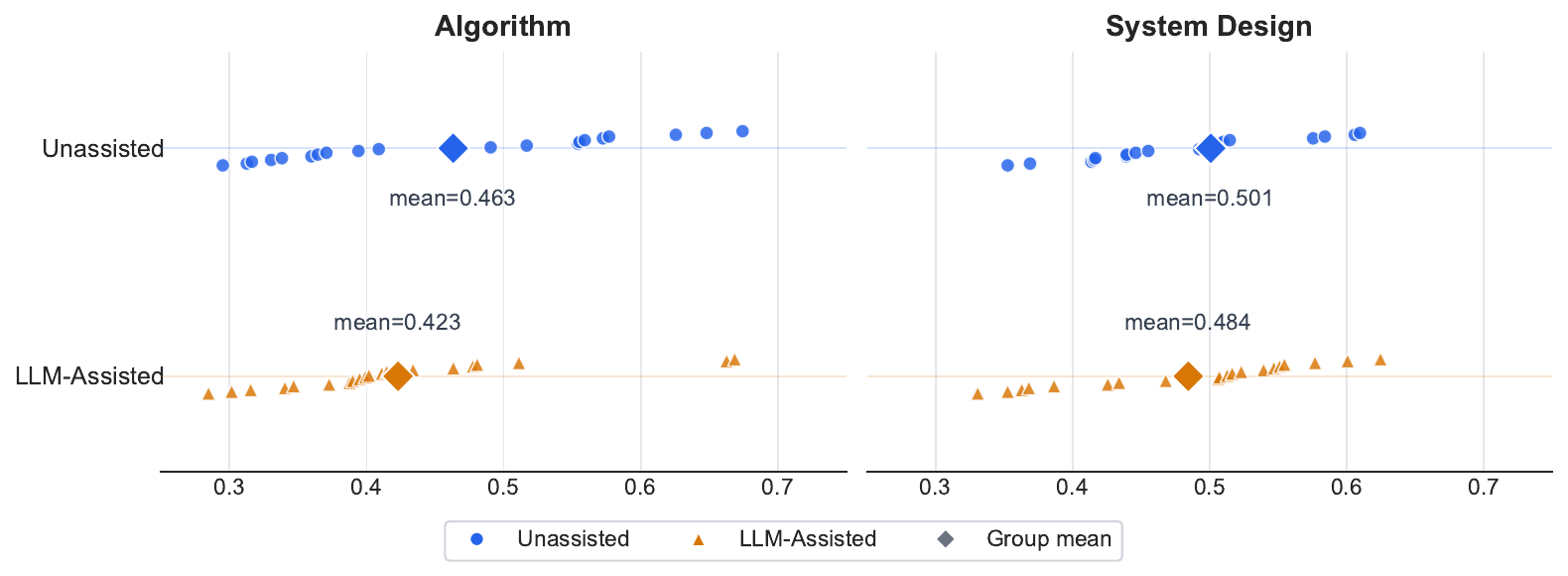}
 \caption{\rrevised{Distance from each solution to the centroid of its own condition, computed with that solution left out of the centroid, for \Atask (left) and \Btask (right). Larger values indicate greater within-condition dispersion; diamonds mark group means. Unassisted solutions are more dispersed in both panels, but the gap is more than twice as large for \Atask{} ($0.463$ vs.\ $0.423$) as for \Btask{} ($0.501$ vs.\ $0.484$).}}
 \Description{Two horizontal strip plots of each solution's distance to the leave-one-out centroid of its own condition, for Algorithmic tasks (left) and System Design tasks (right). Each dot is one solution, blue for Unassisted and orange for LLM-assisted, with diamonds marking group means. For Algorithmic tasks the unassisted mean (0.463) is higher than the LLM-assisted mean (0.423), indicating greater dispersion; for System Design the means are closer (0.501 versus 0.484) and the distributions largely overlap.}
 
 \label{fig:dispersion}
\end{figure*}

\rrevised{We measured, for each solution, its mean cosine distance to the other solutions to the same question in the same condition, and compared conditions with a \emph{linear mixed-effects model fitted on the log scale with crossed random intercepts for participant and question}. The log scale is used because cosine distances are strictly positive and right-skewed, so the treatment term estimates a proportional rather than an additive change; the crossed random intercepts absorb the fact that each participant contributes several solutions and that questions differ in how much room they leave for variation. Significance was assessed by permuting condition labels within each question and refitting the model (1,000 permutations), which accounts both for repeated measures within participants and for the fact that solutions in a condition share pairwise distances. Smaller distances indicate that solutions are more alike, that is, a narrower solution space.}
Figure~\ref{fig:dispersion} shows the corresponding per-solution distances to each condition's own centroid.

\rrevised{Because this result rests on analysis choices that are not forced by the design, we validated it with a set of sensitivity analyses, each re-running the identical model with one choice varied and everything else held fixed. All of them reproduce the findings reported below. The individual analyses, the full model specification, and a table of every specification we ran are given in the supplementary material (``Sensitivity Analysis'').}

\emph{Algorithmic tasks.} \rrevised{LLM-assisted solutions were substantially more similar to one another than Unassisted solutions: mean pairwise distance decreased by $44.9\%$ ($t = -5.45$, permutation $p = 0.005$). Both questions that permit a within-question comparison agree in direction, with mean distance falling from $0.361$ to $0.231$ for A1 and from $0.223$ to $0.112$ for A2. LLM assistance was associated with narrower solution space of algorithmic solutions participants explored. The sensitivity analyses reproduce this narrowing in every specification we ran (permutation $p \le 0.028$ throughout).} The left panel of Figure~\ref{fig:dispersion} shows this as a visible leftward shift of the LLM-assisted distribution.

\emph{System Design tasks.} \rrevised{We found no corresponding effect for \Btask{}: mean pairwise distance was $15.0\%$ higher under LLM assistance, but this difference was not significant ($t = 2.23$, permutation $p = 0.324$), and the four questions disagreed in direction (B2 and B3 more diverse under assistance, B1 and B4 less). The narrowing of the solution space is therefore specific to \Atask{}, where the space of reasonable approaches is comparatively small and LLM suggestions converge on a few canonical algorithms; for open-ended design tasks, participants' solutions remained as varied as those developed without assistance. The sensitivity analyses agree that there is no reliable effect here: no specification we ran reaches significance, and one even reverses the sign of the estimate.} The right panel of Figure~\ref{fig:dispersion} shows the two distributions largely overlapping.

\begin{table}[ht]
\centering
\caption{Distribution of unique ideas across the task types and specific tasks in solutions from unassisted and LLM-assisted conditions.}
\begin{tabular}{|cccl|}
\hline
\textbf{Taskname} & \textbf{Condition} & \textbf{Num of unique ideas} & \textbf{\modified{\% of unique ideas per participant}} \\
\midrule
\multirow{2}{*}{\textbf{Algorithmic}} 
    & Unassisted   & 9  & \CtrlBar{45}  \\
    & LLM-assisted & 6  &  \TreatBar{30} \\
\midrule
\multirow{2}{*}{A1} 
    & Unassisted   & 3  & \CtrlBar{75} \\
    & LLM-assisted & 1  & \TreatBar{50} \\
\multirow{2}{*}{A2} 
    & Unassisted   & 4  & \CtrlBar{50} \\
    & LLM-assisted & 2  & \TreatBar{33} \\
\multirow{2}{*}{A3} 
    & Unassisted   & 3  & \CtrlBar{43} \\
    & LLM-assisted & 1  & \TreatBar{100} \\
\multirow{2}{*}{A4} 
    & Unassisted   & 1  & \CtrlBar{100} \\
    & LLM-assisted & 3  & \TreatBar{27} \\
\midrule
\multirow{2}{*}{\textbf{System Design}} 
    & Unassisted   & 11 & \CtrlBar{55} \\
    & LLM-assisted & 13 &  \TreatBar{65}\\
\midrule
\multirow{2}{*}{B1} 
    & Unassisted   & 3  & \CtrlBar{60} \\
    & LLM-assisted & 2  & \TreatBar{67} \\
\multirow{2}{*}{B2} 
    & Unassisted   & 5  & \CtrlBar{100} \\
    & LLM-assisted & 5  & \TreatBar{100} \\
\multirow{2}{*}{B3} 
    & Unassisted   & 3  & \CtrlBar{70} \\
    & LLM-assisted & 7  & \TreatBar{75} \\
\multirow{2}{*}{B4} 
    & Unassisted   & 3  & \CtrlBar{50} \\
    & LLM-assisted & 1  & \TreatBar{50} \\
\hline
\end{tabular}
\label{tab:bytask}
\end{table}

\subsubsection{Unique Ideas across Conditions.}
\label{sec:rq1:novelty}

To analyze whether LLM-assisted code is more unique, two researchers identified the number of unique ideas in each solution across the two conditions. We found an equal number of unique ideas across solutions in unassisted and LLM-assisted conditions. Table~\ref{tab:bytask} breaks down the distribution of unique ideas across conditions. Both conditions were associated with 18 unique ideas (with no significant difference when compared using the chi-square test of independence).

However, our experimental design involved two types of tasks, and we posit that problems of an algorithmic nature may have less space to explore creative solutions compared to problems about system design. To analyze whether the uniqueness of the solutions differed by task type, we examined ideas across task types (40 \Atask questions and 40 \Btask questions, evenly divided between the two conditions). Table~\ref{tab:bytask} shows the distribution of unique ideas across the task types and specific tasks in solutions from unassisted and LLM-assisted conditions. Although there are no significant differences, we observed some interesting variations. Solutions generated with the help of LLMs have a slightly higher number of unique ideas for system design-type problems (13 compared to 11). Whereas, participants' solutions without LLM assistance for algorithmic problems demonstrated a slightly higher number of distinct ideas (9 vs.\ 6).

We further analyzed the distribution of unique ideas for each task (per participant) in Table~\ref{tab:bytask}. While we observe some task-specific differences, overall trends indicate no significant difference in the distribution of unique ideas across tasks, as determined by a chi-square test. In \Atask{}, participant-generated solutions contained more unique ideas except A3 (queue operations in O(1) complexity). In \Btask{}, solutions generated with the help of LLMs contained more unique ideas for task B1 (a calendar application focusing on scheduling conflicts), while the other tasks showed an equal or slightly lower number of unique ideas compared to participant-generated solutions.

While the count of unique ideas at the individual task level does not differ significantly between conditions, this metric captures only the \textit{number} of distinct strategies, not how \textit{distributed} participants are across those strategies. The embedding-based analysis above complements this finding by showing that even when a similar number of unique approaches exist, LLM-assisted participants cluster more tightly around a few dominant strategies for \Atask{}, whereas unassisted participants spread more evenly across the available solution space.

Our analysis of solution quality and diversity reveals a trade-off between correctness and creative exploration. LLM-assisted solutions were significantly more correct and more often executable (Tables~\ref{tab:correctness} and~\ref{tab:fidelity}), confirming the productivity benefits of code generation tools. However, even after restricting the analysis to correct solutions only, \Atask solutions produced with LLM assistance were substantially more similar to one another (\S\ref{sec:rq1:pairwise}), \rrevised{consistent with participants using LLMs converging, within these sessions, on a narrower set of algorithmic strategies. We found no corresponding effect for \Btask{}.} \rrevised{The complementary analyses reported in the supplementary material point in the same direction: unassisted participants' solutions spread across more of the embedding space and show a higher proportion of distant solution pairs.}

These findings align with and extend prior work on LLM-induced homogenization in creative tasks. While Doshi et al.~\cite{doshi2024generative} demonstrated that LLMs reduce collective diversity in writing, \rrevised{our results show that a comparable effect operates in algorithmic programming, and is not merely a byproduct of LLMs producing fewer errors: it is present even among solutions restricted to correct and partially correct approaches.} To understand how this narrowing occurs during the problem-solving process itself, we turn to RQ2 in the next section.

%% file: Sections/RQ2.tex
\section{RQ2: How does LLM usage affect programmers' creativity in programming steps?}
To evaluate how LLM usage influences the creative process, we analyzed participants' engagement across five programming stages. We compared the frequency and duration of each stage between the Unassisted and LLM-assisted conditions to identify shifts in cognitive effort and creative engagement.

\subsection{Distinct Patterns of Creative Engagement across Programming Stages}

\begin{figure}[ht]
    \centering
    \begin{subfigure}[b]{0.48\linewidth}
        \centering
        \includegraphics[width=\linewidth]{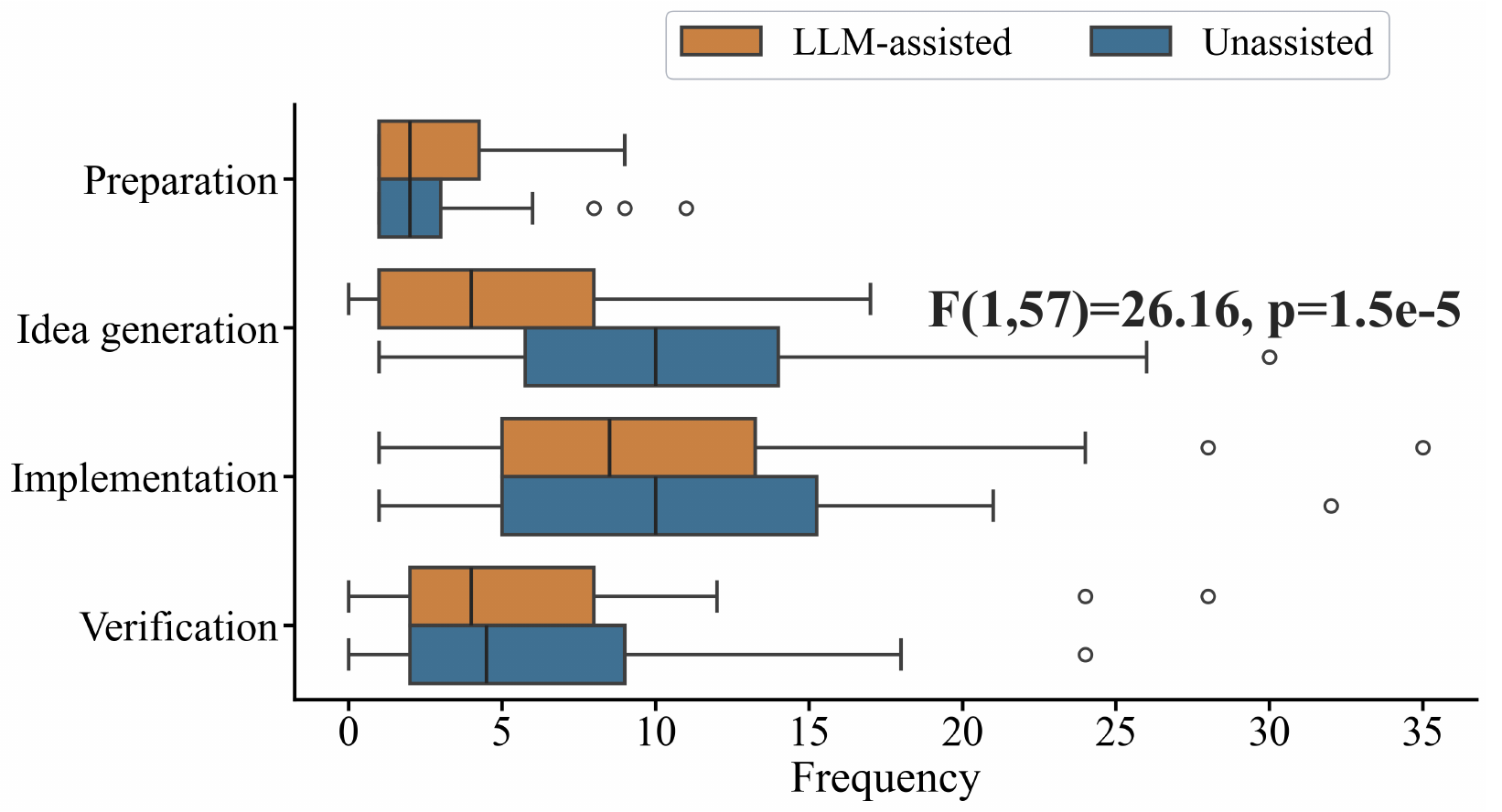}
        \caption{Frequency distribution}
        \label{fig:frequency_condition}
    \end{subfigure}
    \hfill
    \begin{subfigure}[b]{0.48\linewidth}
        \centering
        \includegraphics[width=\linewidth]{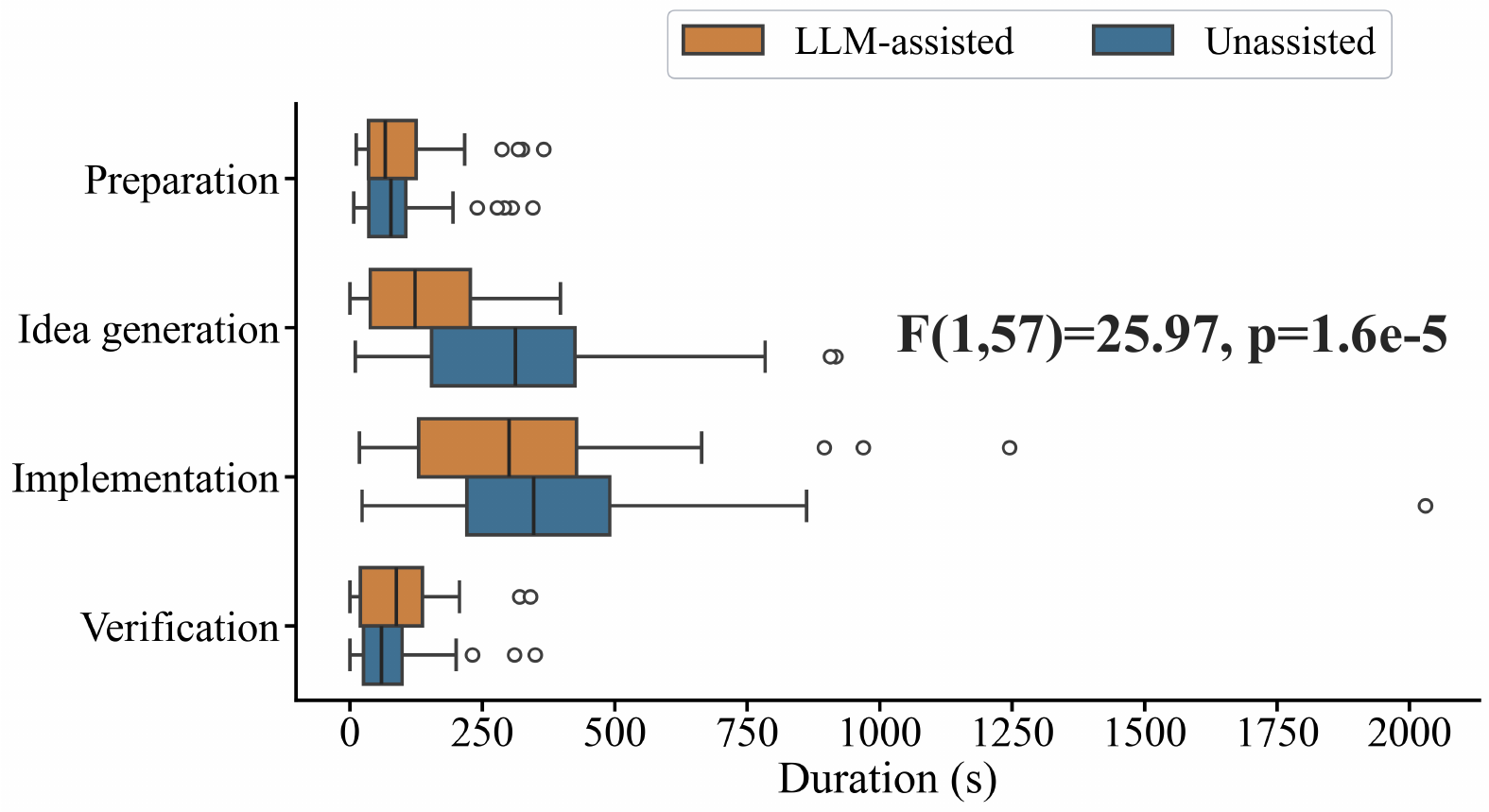}
        \caption{Duration distribution (seconds)}
        \label{fig:duration_condition}
    \end{subfigure}
    \caption{Comparison of frequency and duration distributions of each programming stage in both conditions}

    \Description{Two panels of horizontal box plots across the four programming stages (Preparation, Idea generation, Implementation, Verification), with orange for LLM-assisted and blue for Unassisted. Panel (a) shows the frequency distribution and panel (b) the duration in seconds. At Idea generation the unassisted boxes are shifted toward higher frequency and longer duration, annotated as significant (F(1,57) = 26.16, p = 1.5e-5 for frequency; F(1,57) = 25.97, p = 1.6e-5 for duration); the other stages overlap more.}
    \label{fig:frequency_duration_condition}
\end{figure}

Two researchers quantitatively analyzed the frequency (number of times participants engage in each stage) and duration(length of time participants spent in a stage) of participants' programming stages across 80 tasks, comparing the \textit{Unassisted} and \textit{LLM-assisted} conditions. Figure~\ref{fig:frequency_duration_condition} shows these comparisons: the left panel (Figure~\ref{fig:frequency_condition}) displays the frequency distribution for each stage, while the right panel (Figure~\ref{fig:duration_condition}) displays the duration distribution. As "illumination" is defined as a momentary insight, we did not compare its duration.

To examine whether engagement in each stage differed across the two task types, we compared the frequency and duration of each stage between the unassisted and LLM-assisted conditions. \rrevised{An ART ANOVA test with Holm-Bonferroni correction showed that without LLM access, participants engaged in idea generation significantly more frequently ($F(1, 57) = 26.16$, $p_{\mathrm{holm}} < 0.001$, $\eta_p^2 = 0.32$) and longer ($F(1, 57) = 25.97$, $p_{\mathrm{holm}} < 0.001$, $\eta_p^2 = 0.30$).} Differences in the other stages were not statistically significant, though their distributions appeared notably different. Two researchers conducted Thematic Analysis on participants' interview transcripts to understand the reasons behind these differences. 

\subsubsection{Preparation} 

Participants engaged in preparation more often and for longer durations in the LLM-assisted condition, \rrevised{though this difference was not statistically significant ($F(1, 57) = 0.02$, $p_{\mathrm{holm}} = 1.00$, $\eta_p^2 < 0.01$)}.

In the unassisted condition, preparation was mostly self-directed. All participants engaged in independent problem analysis, with participants relying on internal cognitive strategies to solve requirements. Five participants explicitly drew on past programming experience to frame the current problem, with P23 noting a dynamic programming approach: "let's see if we can use DP to solve this...I would say the number of ways I can make this sum using the first m elements." Three participants (P17, P18, P23) structured their plan by decomposing problems into subtasks first for instance, P23 "wanted to construct a graph based on dependencies and then iterate through it." instead of figuring out the whole solution. 

In contrast, participants in the LLM-assisted condition externalized preparation through dialogue with the LLM assistant. While all participants still engaged in problem understanding, \rrevised{10 participants} (P3, P4, P6, P7, P9 , P12, P13, P16, P20, P21) used the LLM to clarify problem specifications or technical concepts. Some participants used LLMs for better understanding the requirements (P6,P13) or "terminology clarification" (P9, P20) For instance, P6 queried the LLM: "to give [them] an example, so [they] are better able to understand what's a triplet."

Four participants in the LLM-assisted condition (P10, P12, P17, P7) broke problems into steps during preparation and "asked Copilot to first list out all the small tasks of these bigger tasks instead of asking to do everything." (P12). 

\subsubsection{Idea Generation}
\label{sec:ideation}

\rrevised{Idea generation was the one stage that differed significantly between conditions.} Additionally, all participants engaged in idea generation in the unassisted condition, whereas 18 of 20 did so in the LLM-assisted condition.

In the unassisted condition, participants employed internal problem-solving strategies without external assistance. Sixteen participants (P10, P11, P12, P13, P14, P15, P19, P20, P21, P22, P23, P3, P4, P6, P7, P9) generated algorithmic strategies independently, and nine (P10, P13, P16, P17, P23, P3, P5, P6, P7) engaged in brainstorming. Eight participants (P10, P12, P13, P17, P18, P22, P23, P9) drew on prior programming experience with similar data structures or algorithms. Some unassisted participants expressed awareness of the cognitive benefits of unassisted problem-solving.

As P4 reflected:
\blockquote{
Knowing that I had to do it on my own here made me much more capable of thinking because I didn't have that crutch (LLMs) to fall back on.
}

Unassisted participants employed diverse strategies. Some (P11, P17, P20) started with a naive, suboptimal solution and refined it iteratively, whereas others (P3, 6, 10, 19, 21, 22, 23) conducted a breadth-first exploration of multiple algorithms. In contrast, participants with LLM access offloaded this demanding task, either partially or entirely, to the AI. Some [P3, 14, 19, 23] fully delegated the task by prompting the model with the entire problem statement directly. Others used the LLM more selectively as a brainstorming partner. Rather than requesting complete solutions, they collaborated with the AI to refine or validate their own concepts, using it to "narrow down focus" (P13), "divert [their] mind to the correct direction" (P6), or evaluate trade-offs between approaches (P7). 

LLM assistance eases the burden of analyzing the problem space and coming up with feasible solutions \revised{and our results show that it was associated with a significant reduction in both the frequency and duration of the ideation stage} as seen in Figure~\ref{fig:frequency_duration_condition}. However, \revised{delegating ideation to the AI was associated with reduced engagement in the problem-solving process.}
Many participants reported disliking this lack of creative ownership feeling, explaining that they \inlinequote{can't be creative when the AI is good enough} \revised{(P19)}. Yet the convenience of LLMs was highly tempting: even participants who felt fully capable of generating and developing their own ideas still relied heavily on it, especially under the time constraints of the study. P4 described this as a psychological pull toward the "path of least resistance":
\modified{\blockquote{With an LLM, there is a pressure to just give up and ask for an answer, whereas without it, you just go through and do what you have to do. So yeah, it's not a pressure of like \inlinequote{I have to do this.} It's more so \inlinequote{I can}---this could be so much easier. And I could just stop thinking at any moment, and brains are usually like taking the path of least resistance.}}

\subsubsection{Implementation}
\revised{During LLM-assisted tasks, participants spent less time on implementation.} In the unassisted condition, implementation was characterized by direct manual coding, which constituted 99.3\% of implementation activity. LLM-assistance \revised{was associated with a reduction in} this manual coding to  48.2\% of implementation instances. This productivity gain \revised{was associated with} the LLM's ability to help users overcome the syntactic and library-recall hurdles that stalled unassisted participants (\S\ref{sec:rq1:correctness}).


\modified{With the assistance of LLMs, these roadblocks easily vanished, as the AI took over the hassle of translating conceptual ideas into functional code. Eight participants already skipped this stage altogether as they had used the LLM during ideation to generate a complete implementation. The remaining participants leveraged this capability in two ways. Eight participants first outlined a high-level plan or code structure and then relied on the LLM to flesh out most of the working code based on this skeleton.} As P5 described, \inlinequote{After forming my own idea[s] of how the code should work, I gave it (LLM) a prompt on how to incorporate my thinking.} This allowed participants to \inlinequote{get the mundane stuff out of the way}\revised{(P4)} and focus on higher-level design. Four participants used the LLM more like a specialized search engine to address minor issues, such as clarifying \inlinequote{how to define the constructor in Python classes} \revised{(P18)}, or \inlinequote{finding an element/index in an array} ~\revised{(P9, 15, 20)}, etc. This strategy allowed participants to maintain their coding flow without getting bogged down by minor syntax issues.

\subsubsection{Verification}
In the unassisted condition, participants engaged in more frequent but shorter verification stages, whereas participants in the assisted condition completed earlier stages more quickly, leaving them with more time for verification and debugging.

Without AI assistance, verification was a manual and iterative process where participants relied on self-devised, piecemeal checks to ensure their code worked. For example, P14 tested each component of his code sequentially \inlinequote{on some examples to see if it works correctly before moving on}; P6 manually \inlinequote{print everything, tests even small parts of the code repeatedly, so I wouldn’t have to debug it all at once.} \rrevised{Such practice of frequent, low-level testing is consistent with the shorter, more frequent verification cycles observed in the unassisted condition.}

\modified{In the assisted condition, verification strategies depended strongly on who authored the code. Participants who chose to implement the solution themselves \revised{(P10, 15, 16, 17)} all drew on the AI during this stage. They asked the LLM to generate comprehensive test cases to surface overlooked corner cases. Some used it to fix syntax errors or to suggest more optimal solutions ~\revised{(P20)}. Conversely, participants who had the LLM produce most of the implementation shifted their effort toward auditing the AI's work. Of the 16 who delegated implementation to the AI, 11 spent the verification stage manually examining the code quality and ensuring that it actually followed the approach they had in mind. P14, for example, noted that they \inlinequote{never trust[ed] the LLM,} and therefore carefully cross-checked its output on different examples. P3 described a similar skeptical stance toward LLM-generated code: \inlinequote{I... fed my ideas to the LLM to ask it to implement what was in my mind. Then I verified the solutions and debug them.}} \rrevised{Overall, participants who voiced skepticism toward LLM-generated code also described deeper and more deliberate diagnosis, which is consistent with the longer but less frequent verification cycles observed in Figure~\ref{fig:duration_condition}.}

\begin{figure}[b]
    \centering
    \includegraphics[width=\linewidth]{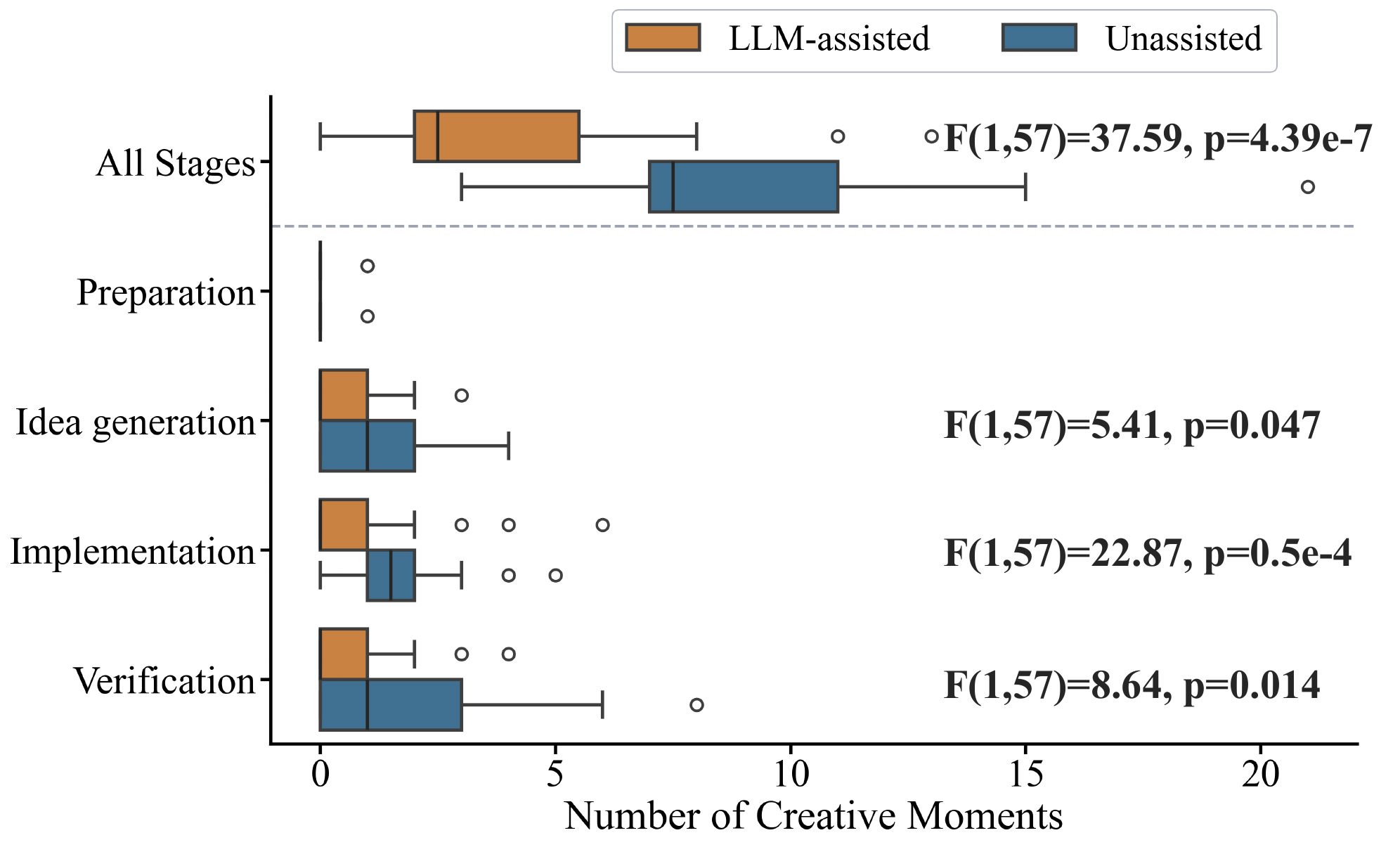}
    \caption{Number of creative moments during each problem solving stage across unassisted and LLM-assisted conditions}
    \Description{Horizontal box plots of the number of creative aha moments per participant, with orange for LLM-assisted and blue for Unassisted. Rows show All Stages followed by Preparation, Idea generation, Implementation, and Verification. Across all stages the unassisted condition shows more creative moments (F(1,57) = 37.59, p = 4.39e-7), with significant differences at Idea generation (p = 0.047), Implementation (p = 0.5e-4), and Verification (p = 0.014); Preparation shows almost none in either condition.}
    \label{fig:aha_moments_steps}
\end{figure}


\subsection{Impact of LLM usage on Creative Moments}

\modified{We compared the number of creative moments (i.e., "aha" moments) experienced by participants across the two conditions within each problem-solving stage. Figure~\ref{fig:aha_moments_steps} shows the boxplot distributions across stages and conditions, \rrevised{along with results from ART ANOVA with Holm-Bonferroni correction}.} Our analysis revealed a significantly lower frequency of creative moments when participants used an LLM assistant \rrevised{($F(1, 57) = 37.59$, $p = 4.39 \times 10^{-7}$, $\eta_p^2 = 0.40$)}. \rrevised{This reduction was consistent across the idea generation ($F(1, 57) = 5.41$, $p_{\mathrm{holm}} = 0.047$, $\eta_p^2 = 0.09$), implementation ($F(1, 57) = 22.87$, $p_{\mathrm{holm}} = 5.05 \times 10^{-5}$, $\eta_p^2 = 0.29$), and verification ($F(1, 57) = 8.64$, $p_{\mathrm{holm}} = 0.014$, $\eta_p^2 = 0.13$) stages.} \rrevised{Notably, this difference was observed alongside the time participants spent in each stage,} \revised{the presence of the LLM was strongly associated with the reduced occurrence of these insights.}

Creative moments were very rare in the initial preparation stage across both conditions. In the unassisted condition, creative moments typically arose from self-guided exploration and iterative sensemaking. Participants experienced breakthroughs when grappling with the problem constraints on their own. For example, insights occurred during the initial framing of the problem, as one participant (P22) realized they had initially misinterpreted the prompt: \inlinequote{I read the question wrongly, initially… it was in the realization of the problem, that I had an aha moment}. Other "aha" or creative moments came from the independent discovery of a viable solution strategy: \inlinequote{I had an 'aha moment' when I figured out a problem could be solved using DP and realized the specific DP relation that was going to work. There were a couple of moments when I was working on a solution and suddenly realized there might be a case where this solution might not work.} ~\revised{(P23)} Furthermore, the challenging process of identifying edge cases and analyzing errors through trial-and-error also sparked many such creative insights.

Conversely, while creative insights still occurred in the LLM-assisted condition, their origin often came from external assistance rather than users' internal discovery. Participants often described moments where the LLM provided a crucial piece of information they had not generated themselves, effectively acting as the catalyst for the insight. For instance, P14 attributed an insight to the model's ability to identify a novel corner case: \inlinequote{...I saw the creativity of LLM, and I liked it, and the way that it dropped the code.} P16 similarly described clarification in preparation through the LLM: \inlinequote{At that moment, I had another interpretation of the question, but by reading the LLM's response, the question actually got clarified for me.}
P6 highlighted the role of LLM in transferring their idea to the actual code: \inlinequote{For the second task, I wasn't getting the answer on my own, so I prompted it. It gave me something that I've read about but didn't know how to implement… now the LLM was giving me the implementation. That's why my trust in it built up… that was an 'aha moment' for me.}

Ultimately, the use of an LLM appears to alter the creative process by offloading the cognitive struggle that often precedes discovery. In the unassisted condition, participants wrestled with ambiguity, debugged errors, and iteratively refined their ideas, processes ripe with opportunities for more "aha" moments. \rrevised{The readily accessible strategies the LLM offered accelerated problem-solving, but co-occurred with fewer instances of the sustained cognitive struggle that preceded insight in the unassisted condition.}

%% file: Sections/RQ3.tex
\section{RQ3: How does LLM support Creativity in LLM-assisted sessions?}
\label{sec:collaborative}
While participants in unassisted conditions experienced more creative moments, using LLMs when programming was still associated with creative moments, albeit fewer. This suggests that LLM use does not eliminate creativity; rather, certain ways of working with the model may be more or less supportive and effective for creative thinking.

\begin{figure}
    \centering
    \includegraphics[width=\linewidth]{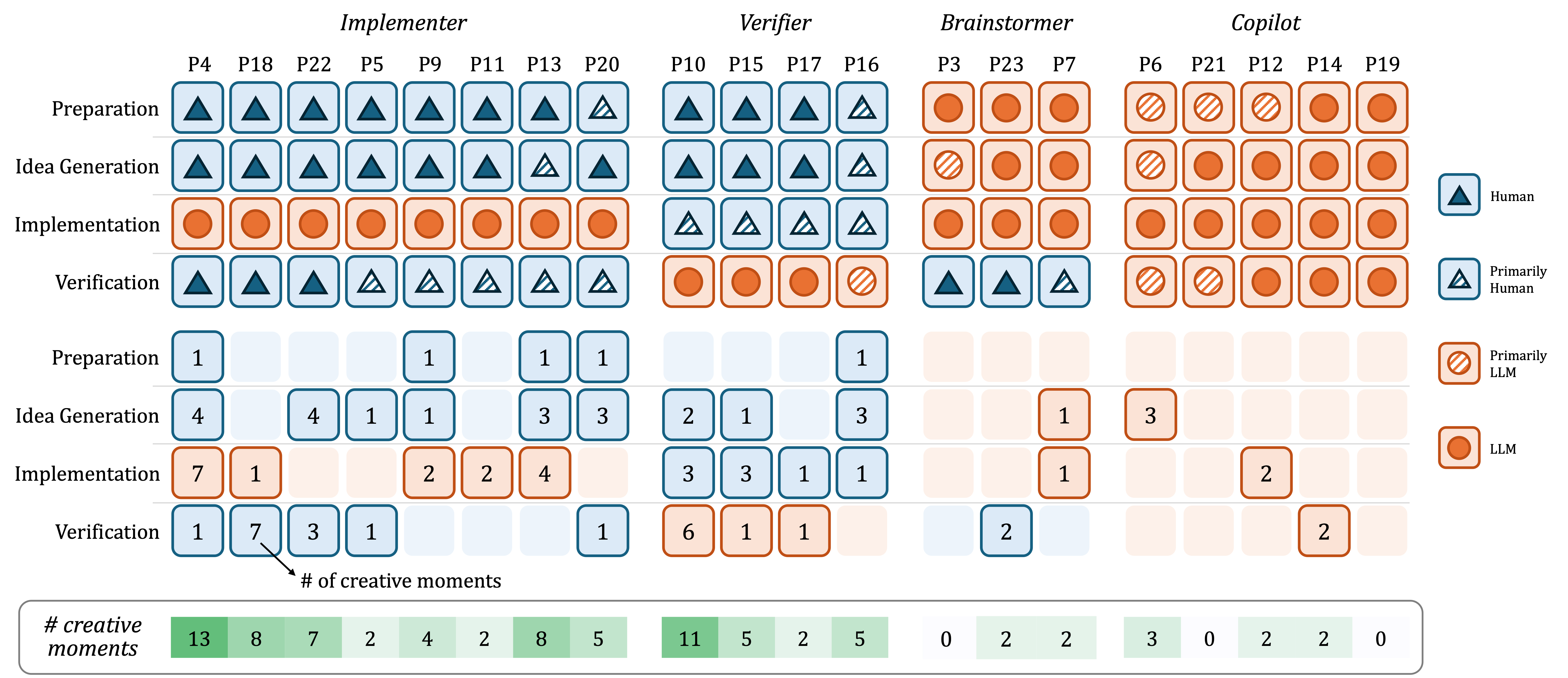}
    \caption{The top grid shows individual participants' collaboration strategies (Human, Primarily Human, Primarily LLM, or Fully LLM) across the four programming stages. The lower grid reports the number of creative moments observed for each participant at each stage. The bottom heatmap summarizes the total number of creative moments per participant.}
    \Description{Grid visualization of the 20 participants grouped into four collaboration models: \CMthree, \CMfour, \CMone, and \CMfour. The top grid places one symbol per participant per stage (Preparation, Idea Generation, Implementation, Verification), with blue symbols for greater human involvement and orange for greater LLM reliance on a four-level scale (Human, Primarily Human, Primarily LLM, LLM). The middle grid gives the count of creative moments per participant per stage, and the bottom row is a green heatmap of each participant's total creative moments, highest for the \CMthree and \CMfour groups and near zero for several \CMfour participants.}
    \label{fig:aha_collaboration}
\end{figure}

To examine this, we analyzed how participants collaborated with the LLM during assisted sessions and how these collaborations shaped both their creative process and their final code. For each participant and each problem-solving stage, we coded the primary agent responsible for the work as one of four types: \textit{Human only}, \textit{Primarily human}, \textit{Primarily LLM}, or \textit{LLM only}.

Figure~\ref{fig:aha_collaboration} summarizes these patterns. The upper panel shows each participant's strategy across the four stages, with blue symbols indicating greater human involvement and orange symbols indicating greater reliance on the LLM. The lower panel reports the number of creative moments observed for each participant across the four stages. Notably, the ways participants collaborated with the LLM clearly fall into four distinct cluster. We refer to these clusters as collaboration models and discuss each of these models in turn below:

\textbf{\CMtwo}. \rrevised{In this most frequent model, eight participants kept creative control and, contrary to the previous pattern, used LLMs only to implement their own ideas.} After spending effort to understand the problem, they would \textit{``iteratively choose different ideas, like, different data structures''}~\revised{(P22)} for the solution\rrevised{, then have LLMs implement the chosen approach based on specific instructions.} \rrevised{This let them focus on thinking while offloading mechanical work. P4, for instance, used the LLM \emph{``to get the mundane stuff out of the way''} so he did not have to \emph{``pause [his] thinking just to do some typing.''}} Two participants \revised{(P5, P18)} also \rrevised{found this model more time-efficient and effective, with P5 noting that the LLM helped them} \emph{``incorporate [their] thinking so the solution is way better than [they] would generate themselves.''} P20 summarizes why participants preferred this model: \blockquote{I don't like using LLMs to do the problem solving for me. I feel like that just makes me constricted to what they can think of and not what I can think of. \rrevised{[\ldots]} I only use them to build the solution to the problem that I came up with.}

\textbf{\CMthree}. Four participants engaged in this pattern of only using LLM to verify and refine solution generated by themselves. This model closely mimics LLM-as-a-judge patterns observed in decision-making~\cite{10.5555/3666122.3668142,li-etal-2025-generation} . Participants mainly used LLM to ``debug their solution'' \revised{(P15,17)} or when they had limited time, and they could not generate a working solution [due to the bug], they would ask \CMfour to debug [their] code (P10). LLM also helped them to verify their code by generating ``[test] cases for [their] current solution.'' \revised{(P16)}


\textbf{\CMone}. Participants following this collaboration mode offloaded the generative thinking to LLMs while keeping close watch and verifying LLM-generated solutions extensively. \rrevised{Three} participants engaged in this collaborative pattern, using LLM to take the lead in preparing, generating a solution idea, and implementing the idea. When preparing, they asked LLM to \textit{``explain the problem''} \revised{(P7)} or \textit{``generating sample input and output''} \revised{(P3)} to better understand the requirements. Then these participants would \rrevised{hand the requirements directly to the LLM, letting it take over} the idea generation and implementation steps. However, they were more involved in the verification step of the process by directly testing the written code and verifying the solution. P23 elaborated this pattern clearly stating: \blockquote{I didn't even have to read through the task, I just gave them to LLM, and then I would look to see whether the output matches what the question actually desired? So I didn't have to think about anything. So, LLM had to do most of the thinking.}

\textbf{\CMfour}. Five participants relied on LLMs throughout their session, collaborating with LLM at each step of problem solving. They \rrevised{had the LLM explain the question when they found it challenging (P21), then} used it to \textit{``list out (breakdown) all the smaller tasks''}~\revised{(P12)} \rrevised{and implement them one by one. This model was more common when participants struggled with a task and fell back on the LLM to avoid high cognitive effort. As P19 put it:} \inlinequote{Having access to LLM made me lazy. Even though I had a creative idea, when I found the problem to be too much effort, I said, no... I didn't want to go there.}

\subsection{\new{Creative moments across collaboration modes}}
\label{sec:aha_cm}

\new{Participants experience distinctly different creative moments across the collaboration modes. Notably, the number of creative moments when participants reserved the more cognitively demanding stages—Preparation and Idea Generation—for themselves is higher. The sparser creative moments in the matrix in Figure~\ref{fig:aha_collaboration} belong to participants who relied on LLMs to take control of ideation.}

\new{
The \textit{\CMtwo} collaboration mode offers a compelling case for LLMs as creativity amplifiers. These participants strategically delegated only the mechanical and routine tasks of implementation to the LLM. This approach \revised{was associated with the highest number of creative insights by} ($\mathbf{49}$ total moments), with a strong concentration in the stages under human control, such as  Idea Generation ($\mathbf{16}$ moments). As P4 stated, the goal was to \textit{``get the mundane stuff out of the way''} to keep \textit{``thinking more abstract.''} By reducing the cognitive load of syntax and typing, participants were free to explore multiple solution pathways and data structures, preserving the very processes that spark innovative insights.
\CMthree mode represents participants using LLMs for refinement. These participants primarily used the LLM as a debugging tool or quality assurance tool to verify and refine their own existing solutions. Creative insights were understandably focused later in the process, with eight moments in the Implementation stage and eight in the Verification stage, as participants wrestled with complex code they had written themselves. }

\new{\rrevised{In contrast, the two collaboration modes in which participants offloaded generation to the LLM (fully or partially) recorded markedly fewer creative moments. Participants in the \textit{\CMone} and \textit{\CMfour} modes described delegating the generative work rather than undertaking it themselves.} The \textit{\CMone} mode, where the LLM took the lead on generation and implementation, resulted in only four total creative moments across all stages for the three participants combined. The participant's sole focus was external verification, explicitly stating, \textit{``I didn't have to think about anything.''} Even more telling, the \textit{\CMfour} collaboration mode, which relied on the LLM throughout all steps, recorded \revised{three} creative moments during the crucial idea generation stage, and \rrevised{seven} moments overall. This lack of engagement aligns with P19’s confession: having access to the LLM made them \textit{``lazy,''} choosing to avoid effort even when a creative idea was present.}

\subsection{Characteristics of Solution across Collaboration Modes}



\rrevised{Table~\ref{collabmodecode_metrics_kruskal} shows the average of code characteristic codes of each collaboration mode. A Kruskal-Wallis test revealed that the Number of function calls, lines of code, logical complexity, and code maintainability scores are significantly different across different categories. For the collaboration-mode analysis, modes were assigned at the task-solution level rather than the participant level. Because each participant completed two LLM-assisted tasks, the same participant could contribute two solutions with different collaboration modes. Therefore, collaboration-mode counts refer to assisted solutions, not participants.}


\textit{Implementers} produced solutions with the highest average lines of code ($\mathbf{81.91}$) and the highest average Logical Complexity ($\mathbf{17.91}$) (See table~\ref{collabmodecode_metrics_kruskal}). Furthermore, this mode generated the highest ratio of Unique Ideas ($\mathbf{0.72}$), reflecting the richness and originality of the human-driven conceptual phase. \rrevised{This pattern is consistent with the possibility that reserving ideation for the human while delegating implementation supports a richer conceptual phase.}

Intriguingly, \textit{Verifiers} produced the code with the highest average Lines of Code ($\mathbf{104.75}$) and the highest Logical Complexity ($\mathbf{19.92}$). This suggests that participants reserved the ``LLM-as-the-judge'' role for their most intricate, human-generated solutions that were difficult to debug manually. 


\rrevised{The \textit{\CMone} and \textit{\CMfour} modes also differed from the others in the structural characteristics of the code produced.} The \textit{\CMone} mode produced solutions with significantly lower logical complexity of $\mathbf{3.71}$ and fewer lines of code ($\mathbf{33.86}$). The \textit{\CMfour} mode produced the second lowest across these two metrics. They also generated a lower number of Unique Ideas ($\mathbf{0.42}$ for \CMone and $\mathbf{0.66}$ for \CMfour). \rrevised{While the code might be functionally correct, it was structurally simpler; we report this as a co-occurring characteristic rather than as evidence of low effort, and note that fully offloading ideation was associated with structurally simpler solutions and fewer creative moments.}

\begin{table}[ht]
\centering
\caption{Kruskal-Wallis Comparisons of Code Metrics Across Collaboration Modes}
\resizebox{\textwidth}{!}{
\begin{tabular}{lcccccc}
\hline
\textbf{Metric} &
\textbf{\makecell{Mean\\(\CMone)}} &
\textbf{\makecell{Mean\\(\CMtwo)}} &
\textbf{\makecell{Mean\\(\CMthree)}} &
\textbf{\makecell{Mean\\(\CMfour)}} &
\textbf{H-stat} &
\textbf{P-value} \\
\hline

Number of Function Calls & 
6.29 & 18.27 & 22.92 & 14.83 &
15.3946 & \textbf{\cellcolor{yellow!25}0.00151} \\

Number of External Libraries Used & 
0.00 & 0.36 & 0.17 & 0.17 &
3.6834 & 0.29775 \\

Lines of Code &
33.86 & 81.91 & 104.75 & 55.50 &
16.2149 & \textbf{\cellcolor{yellow!25}0.00102} \\

Logical Complexity &
3.71 & 17.91 & 19.92 & 14.67 &
12.2862 & \textbf{\cellcolor{yellow!25}0.00646} \\

Code Maintainability &
87.97 & 74.59 & 68.33 & 79.86 &
15.9640 & \textbf{\cellcolor{yellow!25}0.00115} \\

Use of Comprehensions (List/Dict/Set) &
0.00 & 1.00 & 1.17 & 0.00 &
7.3938 & 0.06035 \\

Recursive Functions Used &
1.14 & 1.82 & 1.00 & 1.00 &
2.4289 & 0.48828 \\

Average Function Length (Lines per Function) &
7.50 & 10.39 & 14.15 & 9.95 &
1.9831 & 0.57591 \\

\hline
\end{tabular}}
\label{collabmodecode_metrics_kruskal}
\end{table}

\rrevised{What distinguished sessions with more creative moments was not how often the LLM was used, but which stages participants kept for themselves. The \CMtwo pattern, in which participants retained ideation and delegated implementation, co-occurred with the most creative moments in our sample, while fully offloading generative thinking (\CMone) or relying on the LLM throughout (\CMfour) co-occurred with the fewest. We frame this as a hypothesis for future experimental work.}


%% file: Sections/Discussion.tex
\section{Implications}
\subsection{Implications for Individual Developers}

Participants described a powerful psychological pull toward delegating thinking to the LLM, even when they were fully capable of generating ideas themselves (P19). \rrevised{The presence of an AI as a ``safety net'' (P23) accompanied a shift in how participants engaged with problems (\S\ref{sec:ideation}).} The practical implication is that individual developers should cultivate deliberate habits around when to consult an LLM.

\rrevised{The Implementer collaboration mode co-occurred with the highest raw count of creative moments (49 total) and the highest ratio of unique ideas. By contrast, the Copilot and Brainstormer modes, in which participants offloaded ideation to the AI, co-occurred with the fewest creative moments (7 and 4, respectively) and structurally simpler code.}

\rrevised{For individual developers, this suggests a heuristic worth testing: treating LLMs as implementation accelerators rather than idea generators. Using AI to handle syntactic boilerplate, library lookups, and routine code scaffolding may free cognitive bandwidth for higher-order creative work, such as solution architecture and algorithm selection, while preserving the cognitive struggle our participants associated with novel insights. We present this as a hypothesis for future experimental work rather than as validated guidance.}

\subsection{Implications for LLM-based tool builders}

\rrevised{Our findings point to a tension: the same LLM capability that produced more correct and more complete solutions co-occurred with shorter ideation and fewer reported insights.} This tension between productivity and creativity challenges established assumptions about how generative AI assistance should be designed to support programming. We discuss strategies for designing intelligent creativity support tools.

Creativity support tools (CSTs) have traditionally focused on enhancing user capabilities while preserving creative agency~\cite{cherry2014quantifying, kantosalo2019mapping}. The integration of LLMs introduces new complexities that necessitate a reconsideration of core design principles. 
The four distinct collaboration models from our findings reveal that the design of AI systems influences creative outcomes, suggesting that one-size-fits-all approaches to AI assistance may be inadequate for supporting diverse creative practices.
We envisage two design directives for LLM-based creativity support tools that address the creativity-productivity tension:
\emph{Intentional Friction to Increase Creativity.}
Participants using LLMs experienced fewer creative moments while producing more
correct solutions, and the insights they did report clustered in longer, more
effortful idea generation phases. This sits uneasily with a long-standing
assumption in creativity support tools, expressed most clearly in Shneiderman's
framework of ``collecting, relating, creating, and donating''~\cite{shneiderman2002creativity,shneiderman2007creativity},
that lowering technical barriers and reducing repetitive effort enhances
creativity~\cite{fila2014m, oladele2025future}. \rrevised{Our results are
consistent with the possibility that some effortful engagement instead supports
insight, and they raise the question of whether tools optimized solely for speed
and correctness undermine the processes that produce it.}

\rrevised{A promising direction for future creativity support tools, then, is to preserve certain effort deliberately rather than remove it.} One form this could
take is for models to honor the scope of a prompt: asked for ideas, the system
returns ideas, not an implementation. Another is to decompose problem solving
into discrete, human-approved stages rather than resolving it in a single pass,
for example \inlinequote{here is an approach; shall I proceed to implementation?},
creating checkpoints at which creative agency is exercised rather than bypassed.
This follows the principle of \textit{scalable oversight}~\cite{Kenton2024On,Kim2025The,He2025Plan-Then-Execute},
which argues for keeping humans meaningfully in the loop as AI capabilities grow.
Code generation tools such as Cursor partially adopt confirmation gating, but
still cast the user as verifier rather than as a participant in planning.
\rrevised{Whether such friction increases creative output or simply slows users
down is an open question our study does not answer, and one we see as a natural
next step.}

\emph{Surfacing Creative Provenance.}
A recurring frustration participants described was that LLMs frequently exceeded
their prompts: even when asked for brainstorming help or pseudocode, the model
returned a complete solution. Several found this completeness discouraging,
explaining that reshaping a near-finished solution into their own vision felt
harder than simply accepting it. \rrevised{Participants also described losing
track of where an idea had come from over the course of a session, beginning
with a clear conceptual direction and arriving somewhere else without noticing
the point at which it diverged.}

\rrevised{A second direction for future tools, then, is to make the origin of
ideas visible rather than leaving it implicit.} We sketch a \textit{creative
provenance} layer embedded in the interaction interface, maintaining a shared
record of how each idea was introduced, transformed, or discarded during
collaboration, and flagging the points at which the model reframes a user's idea,
narrows its scope, or substitutes an alternative direction. Extending
interpretability work on chain-of-thought reasoning~\cite{10.5555/3600270.3602070},
such a layer would expose the reasoning trajectory rather than only the model's
outputs, allowing users to recognize creative drift and return to an earlier
state where their intent was still intact. \rrevised{We have not implemented or
evaluated this design. Beyond its use in tools, we see a record of how user
intent expands or narrows across a session as a useful artifact for future
research on creative ownership in human-AI collaboration.}

\subsection{\rrevised{Implications for the Developer Community}}

\rrevised{Beyond individual developers and tool builders, our findings raise
questions for the programming profession at large. We offer these as conjectures
that our design cannot test directly, but that we believe warrant attention.}

\rrevised{Within our sample, algorithmic solutions produced with LLM assistance
were more similar to one another than those produced without it. LLMs are
trained on large repositories of existing code and are
biased toward statistically common solutions. When many developers consult
similar systems on similar problems, this convergence may aggregate into a
narrowing of the solution space that no individual developer would notice.}

\rrevised{A second tension concerns maintenance. LLM assistance produced code
that was more correct and complete but scored lower on maintainability and was
structurally more complex. If that pattern holds at scale, a widespread shift
toward code optimized for correctness could accumulate technical debt, shifting
cost from the moment of writing to the teams who later extend and maintain the
work.}

\rrevised{Finally, these dynamics interact with how developer contribution is
valued. Practices that treat code output volume as the primary measure of
productivity may reward precisely the heavy-delegation patterns that, in our
sessions, co-occurred with the fewest reported insights, and community-level
responses such as norms for disclosing AI contributions may be worth exploring.
Establishing whether any of this holds in practice requires longitudinal and
field studies. The `path of least resistance' we observed is an individual
behavior, but it may not be only an individual one.}

%% file: Sections/Limitations.tex
\section{Limitations}

While our study provides valuable insights into LLM-assisted creativity in programming tasks, several limitations should be acknowledged when interpreting these findings.

\textbf{Study Scope and Generalizability:} Our investigation was constrained to Python programming tasks using only GitHub Copilot as the LLM-assistant tool. While Python represents one of the most widely adopted programming languages and Copilot is among the most prevalent LLM coding assistants, this focus may limit the generalizability of our findings to other programming languages or AI tools. However, our recruitment strategy successfully captured participants with diverse experience levels across both Python and Copilot usage and our findings account for diverse programming and LLM usage backgrounds. We examined creativity through two distinct task types, algorithmic and system design problems. While these were chosen to reflect common scenarios in software development and coding interviews, they do not encompass the full spectrum of real-world programming activities. To ensure our findings were not task-specific, we designed multiple distinct problems within each category. Nevertheless, our experimental design prioritized balancing conditions across these two high-level types rather than controlling for finer-grained attributes (e.g., difficulty, specific problem constraints) within them. \revised{Additionally, our tasks were short, single-session exercises which may limit generalizability to longer-horizon engineering work.}
Consequently, our conclusions are most robust when generalized to the broad distinction between algorithmic and system design creativity.
\textbf{External Validity:} Our findings are grounded in programming tasks, which represent a specific subset of creative problem-solving activities that involve computational and logical thinking. While programming tasks offer valuable insights into human-AI collaboration in structured creative domains, they may not fully capture the nuances of creativity in less structured domains such as artistic creation, open-ended ideation, or narrative development. Future research should investigate LLM impact on creativity across diverse task characteristics and domains to establish broader theoretical foundations.
\textbf{Sample Size:} Our study included 20 participants, resulting in 40 task instances per experimental condition. While these numbers align with sample sizes commonly employed in prior HCI research examining creativity and human-AI interaction, they represent a moderate sample size that may limit the statistical power for detecting smaller effect sizes or interaction effects between variables. \new{Future work should replicate and extend this study with professional developers and in a less controlled environment, whose domain expertise, workflows, and interaction patterns with AI tools may differ substantially.}

\textbf{Construct Validity:} To strengthen construct validity, we employed expert evaluation protocols, triangulated findings with automated code metrics. However, the multifaceted nature of creativity means that our measures may not fully capture all aspects of creative thinking and output.

\textbf{Internal Validity:} To maintain internal validity, we employed multiple expert raters for qualitative coding steps. Our coding process included inter-rater reliability assessments and expert consensus protocols to ensure consistent evaluation standards.

%% file: Sections/Conclusion.tex
\section{Conclusion}

This study examined how Large Language Models impact creativity in programming tasks, investigating both the cognitive processes underlying creative problem-solving and the characteristics of creative solutions when humans collaborate with LLMs. Through a within-subject experiment with 20 participants across Unassisted and LLM-assisted conditions, we explored the nuanced relationship between LLM assistance and human creativity.

Our findings reveal two critical insights into human-LLM creative collaboration. \rrevised{First, while LLM assistance accelerated problem-solving and provided valuable cognitive scaffolding, participants in the LLM-assisted condition reported fewer insight moments and described those they did have as prompted by the model rather than reached independently.} \rrevised{Second, the collaboration modes participants adopted varied in how many creative moments they co-occurred with. The mode associated with the most creative moments leveraged LLMs for routine implementation while participants retained the conceptual and ideation phases.}

Based on our findings, we explored design opportunities for creativity support tool builders that align with creativity in problem-solving, rather than the current tools that primarily focus on the idea-generation step of creativity. We investigate ways to manage the trade-off between mental effort and creative outcomes by introducing adaptive, targeted support and outlining critical design guidelines for creating LLMs that encourage creativity.